\documentclass[twoside,twocolumn,9pt]{article}
\usepackage{etoolbox}

\usepackage{extsizes}
\usepackage[super,sort&compress,comma]{natbib} 
\usepackage[version=3]{mhchem}
\usepackage[left=1.5cm, right=1.5cm, top=1.785cm, bottom=2.0cm]{geometry}
\usepackage{balance}
\usepackage{mathptmx}
\usepackage{sectsty}
\usepackage{graphicx} 
\usepackage{lastpage}
\usepackage[format=plain,justification=justified,singlelinecheck=false,font={stretch=1.125,small,sf},labelfont=bf,labelsep=space]{caption}
\usepackage{float}
\usepackage{fancyhdr}
\usepackage{fnpos}
\usepackage[english]{babel}
\addto{\captionsenglish}{%
  \renewcommand{\refname}{Notes and references}
}
\usepackage{array}
\usepackage{droidsans}
\usepackage{charter}
\usepackage[T1]{fontenc}
\usepackage[usenames,dvipsnames]{xcolor}
\usepackage{setspace}
\usepackage[compact]{titlesec}
\usepackage{hyperref}
\usepackage{amssymb,amsmath}
\usepackage{ulem}
\usepackage{epstopdf}
\definecolor{cream}{RGB}{222,217,201}
\newcommand*\Laplace{\mathop{}\!\mathbin\bigtriangleup}

\fancypagestyle{plain}{
\renewcommand{\headrulewidth}{0pt}
}
\begin{document}
\makeFNbottom
\makeatletter
\renewcommand\LARGE{\@setfontsize\LARGE{15pt}{17}}
\renewcommand\Large{\@setfontsize\Large{12pt}{14}}
\renewcommand\large{\@setfontsize\large{10pt}{12}}
\renewcommand\footnotesize{\@setfontsize\footnotesize{7pt}{10}}
\makeatother

\renewcommand{\thefootnote}{\fnsymbol{footnote}}
\renewcommand\footnoterule{\vspace*{1pt}%
\color{cream}\hrule width 3.5in height 0.4pt \color{black}\vspace*{5pt}} 
\setcounter{secnumdepth}{5}

\makeatletter 
\renewcommand\@biblabel[1]{#1}            
\renewcommand\@makefntext[1]%
{\noindent\makebox[0pt][r]{\@thefnmark\,}#1}
\makeatother 
\renewcommand{\figurename}{\small{Fig.}~}
\sectionfont{\sffamily\Large}
\subsectionfont{\normalsize}
\subsubsectionfont{\bf}
\setstretch{1.125} 
\setlength{\skip\footins}{0.8cm}
\setlength{\footnotesep}{0.25cm}
\setlength{\jot}{10pt}
\titlespacing*{\section}{0pt}{4pt}{4pt}
\titlespacing*{\subsection}{0pt}{15pt}{1pt}

\fancyfoot{}
\fancyfoot[LO,RE]{\vspace{-7.1pt}\includegraphics[height=9pt]{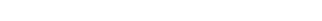}}
\fancyfoot[CO]{\vspace{-7.1pt}\hspace{13.2cm}\includegraphics{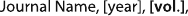}}
\fancyfoot[CE]{\vspace{-7.2pt}\hspace{-14.2cm}\includegraphics{head_foot/RF}}
\fancyfoot[RO]{\footnotesize{\sffamily{1--\pageref{LastPage} ~\textbar  \hspace{2pt}\thepage}}}
\fancyfoot[LE]{\footnotesize{\sffamily{\thepage~\textbar\hspace{3.45cm} 1--\pageref{LastPage}}}}
\fancyhead{}
\renewcommand{\headrulewidth}{0pt} 
\renewcommand{\footrulewidth}{0pt}
\setlength{\arrayrulewidth}{1pt}
\setlength{\columnsep}{6.5mm}
\setlength\bibsep{1pt}

\makeatletter 
\newlength{\figrulesep} 
\setlength{\figrulesep}{0.5\textfloatsep} 

\newcommand{\topfigrule}{\vspace*{-1pt}%
\noindent{\color{cream}\rule[-\figrulesep]{\columnwidth}{1.5pt}} }

\newcommand{\botfigrule}{\vspace*{-2pt}%
\noindent{\color{cream}\rule[\figrulesep]{\columnwidth}{1.5pt}} }

\newcommand{\dblfigrule}{\vspace*{-1pt}%
\noindent{\color{cream}\rule[-\figrulesep]{\textwidth}{1.5pt}} }

\makeatother

\twocolumn[
  \begin{@twocolumnfalse}
{\includegraphics[height=30pt]{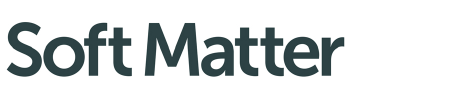}\hfill\raisebox{0pt}[0pt][0pt]{\includegraphics[height=55pt]{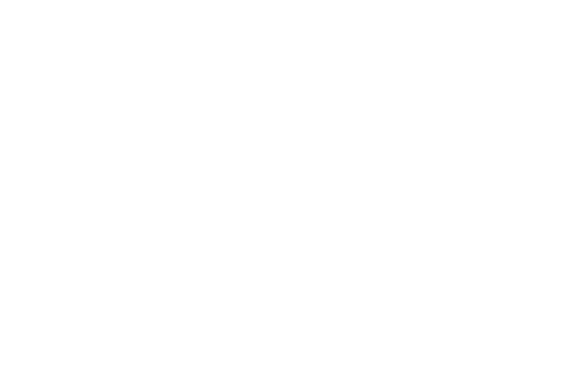}}\\[1ex]
\includegraphics[width=18.5cm]{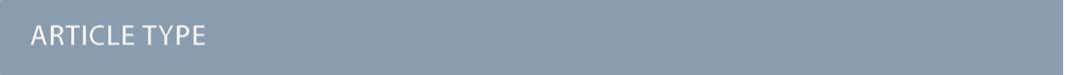}}\par
\vspace{1em}
\sffamily
\begin{tabular}{m{4.5cm} p{13.5cm} }

\includegraphics{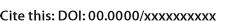} & \noindent\LARGE{\textbf{Turing patterns on polymerized membranes:  a coarse-grained lattice modelling with internal degree of freedom for polymer direction$^\dag$}} \\
\vspace{0.3cm} & \vspace{0.3cm} \\
 & \noindent\large{
  	Fumitake Kato,\textit{$^{a}$}
    Hiroshi Koibuchi,\textit{$^{a}$}\textit{$^{\ast}$}
 	Elie Bretin,\textit{$^{b}$}
 	Camille Carvalho,\textit{$^{b}$}
 	Roland Denis,\textit{$^{b}$}
 	Simon Masnou,\textit{$^{b}$}
 	Madoka Nakayama,\textit{$^{c}$}
 	Sohei Tasaki,\textit{$^{d}$} and Tetsuya Uchimoto\textit{$^{e,}$}\textit{$^{f}$}} 
\\

\includegraphics{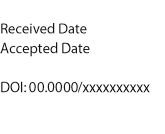} & \noindent\normalsize{We numerically study Turing patterns (TPs) on two-dimensional surfaces with a square boundary in ${\bf R}^3$ using a surface model for polymerized membranes. The variables used to describe the membranes correspond to two distinct degrees of freedom:  an internal degree of freedom for the polymer directions in addition to the positional degree of freedom. This generalised surface model enables us to identify a non-trivial interference between the TP system and the membranes. To this end,  we employ a hybrid numerical technique, utilising Monte Carlo updates for membrane configurations and discrete time iterations for the FitzHugh-Nagumo type Turing equation. The simulation results clearly show that anisotropies in the mechanical deformation properties, particularly the easy axes associated with the stretching and bending of the membranes, determine the direction of the TPs to be perpendicular or parallel to the easy axes. Additionally, by calculating the dependence of the maximum entropy on the internal degree of freedom, we can obtain information on the relaxation with respect to the polymer structure. This crucial information serves to remind us that non-equilibrium configurations can be captured within the canonical Monte Carlo simulations.
}
\end{tabular}

 \end{@twocolumnfalse} \vspace{0.6cm}

  ]

\renewcommand*\rmdefault{bch}\normalfont\upshape
\rmfamily
\section*{}
\vspace{-1cm}


\footnotetext{\textit{$^{a}$ National Institute of Technology (KOSEN), Ibaraki College, Hitachinaka, Japan, E-mail: katobnbf@ibaraki-ct.ac.jp, koi-hiro@sendai-nct.ac.jp}}
\footnotetext{\textit{$^{b}$ Institut Camille Jordan (ICJ), CNRS, INSA, Universite Claude Bernard Lyon 1, Villeurbanne, France, E-mail: elie.bretin@insa-lyon.fr, camille.carvalho@insa-lyon.fr, denis@math.univ-lyon1.fr, masnou@math.univ-lyon1.fr}}
\footnotetext{\textit{$^{c}$ Institute for Liberal Arts, Institute of Science Tokyo, Japan, Email: nakayama.madoka@tmd.ac.jp}}
\footnotetext{\textit{$^{d}$ Department of Mathematics, Hokkaido University, Sapporo, Japan, E-mail: tasaki@math.sci.hokudai.ac.jp}}
\footnotetext{\textit{$^{e}$ Institute of Fluid Science (IFS), Tohoku University, Sendai, Japan, E-mail: uchimoto@tohoku.ac.jp }}
\footnotetext{\textit{$^{f}$ ELyTMaX, CNRS-Universite de Lyon-Tohoku University, Sendai, Japan. }}

\footnotetext{\dag~Electronic Supplementary Information (ESI) available: [details of any supplementary information available should be included here]. See DOI: 10.1039/cXsm00000x/}




\section{Introduction \label{Sec:intro}}





The Turing patterns (TPs), which manifest as spatially periodic and macroscopic patterns on zebras and fish, are described by a pair of partial differential equations. These are known as the reaction-diffusion (RD)  equations, consisting of reaction and diffusion terms  \cite{FitzHugh-BP1961,Nagumo-etal-ProcIRE1962,Gierer-Meinhardt-Kyber1972,Koch-Meinhardt-RMP1994}.   TPs can be observed microscopically in materials, as evidenced by studies such as those conducted by Tan et al. \cite{Tan-etal-Science2018} and Fuseya et al \cite{Fuseya-etal-NatPhys2021}.  In addition,  it is reasonable to extend the discrete Laplacian to the network Laplacian, and the networks are studied in the context of the physics and mathematics developed for TPs \cite{Othmer-Scriven-JTB1971,Nakao-etal-NatPhys2010,Asllani-etal-Ncom2014,Carletti-Nakao-PRE2020,Asllani-etal-NatCom2014,Asllani-etal-PRE2014,Petit-etal-PhysA2016}.  
Furthermore, the TP system has been studied as an inverse problem using the machine learning technique of physics-informed neural networks  to identify the parameter sets \cite{Schnorr-ML2023,Giampaolo-etal-HAL04456081}.

In the RD equation with the reaction terms $f$ and $g$ given by
\begin{eqnarray}
		\label{RD-eq-Eucl}
		\frac {\partial u}{\partial t}=D_u \Laplace u +f(u ,v),  \quad
	\frac {\partial v}{\partial t}=D_v  \Laplace v+ g(u,v), 
\end{eqnarray}
the variables $u$ and $v$ are typically referred to as activators and inhibitors, respectively, representing chemical reactants.  These equations represent an interference between the two variables (Appendix \ref{App-A}), with a significant difference in the diffusion coefficients $D_u$ and $D_v$ \cite{Kondo-Nature1995,Shoji-etal-DevDyn2003,Iwamoto-Shoji-RIMS2018}. This interference is the underlying cause of the emergence of the Turing instability, which is responsible for the formation of TPs \cite{Sekimura-etal-PRSL2000,Kondo-Miura-Science2010,Bullara-Decker-NatCom2015,Vittadello-etal-PTransA2021}.

It is important to note that the Turing pattern is classified within the context of a non-equilibrium steady state configuration. This is due to the fact that the interaction terms in Eq. (\ref{RD-eq-Eucl}), specifically $f(u,v)$ and $g(u,v)$,  are nonlinear and mathematically represent  "chemical reactions" for $u$ and $v$ \cite{note-1}. Chemical reactions are a typical process that occurs in non-equilibrium states. Nevertheless, the patterns are physically understood to represent the minimum energy state of a certain Hamiltonian for the TP system $(u,v)$.

The interference between $u$ and $v$ is reflected in the Hamiltonians $H_{u}$ and $AH_{v}$, which correspond to the steady-state RD equations for $u$ and $v$, respectively. In this context, the sign of $H_{u}$ is opposite to that of $AH_{v}$, where $A(<\!0)$ is a negative constant \cite{Freitas-Rocha-JDE2001} (Appendix \ref{App-A}). As a consequence of this property, the stationary solutions  $(u,v)$ of the RD equations satisfy the stability condition of $H\!=\!H_u\!+\!AH_v$ such that  $\delta H\!=\!\delta (H_u\!+\!A H_v)\!=\!0$ even under $\delta H_{u}^D\!\not=\!0$ and $\delta H_{v}^D\!\not=\!0$, where $H_u(u,v)\!=\!D_uH_u^D\!+\!H_u^R$ and $H_v(u,v)\!=\!D_vH_v^D\!+\!H_v^R$ with the diffusion and reaction Hamiltonians $H_{u,v}^D$ and $H_{u,v}^R$ (Appendices \ref{App-B},\ref{App-C}). This non-trivial interaction provided by the terms $H_u^R$ and $AH_v^R$ gives rise to a steady state solution of the RD equation, corresponding to  $\delta H\!=\!0$, that differs from the minimum energy states of the individual diffusion systems $u$ and $v$ corresponding to $\delta H_{u}^D\!=\!\delta H_{v}^D\!=\!0$.

In addition to these non-trivial interactions in the TP system $(u,v)$, the direction dependence in the $u$ variation, which is caused by the anisotropic diffusion of $u$, influences the anisotropy in the $v$ variation, and consequently, anisotropic TPs arise \cite{Shoji-etal-DevDyn2003,Iwamoto-Shoji-RIMS2018}. Therefore, it would be interesting to combine the diffusion systems $u$ and $v$ with a reaction system  \cite{Castets-etal-PRL1990,Chiu-Chaim-PRE2008}, as well as to combine the TP system with a new set of variables \cite{Dziekan-etal-JCP2012}.

Such an extended TP system is studied in Ref. \cite{Vandin-etal-Softmatt2016} on membranes. TPs on dynamically triangulated surfaces are studied to clarify the role of the reaction wave on cell membranes \cite{Peleg-etal-PlosOne2011,Wu-etal-NatCom2018,Tamemoto-Noguchi-SciRep2020,Tamemoto-Noguchi-SM2021,Tamemoto-Noguchi-PRE2022,Noguchi-SciRep2023}. The activator $u$ is regarded as a concentration of proteins that are coupled to membrane deformation. The combined systems are simulated to gain insight into the shape deformation of membranes in the presence of proteins and the reaction-diffusion waves propagating on membranes \cite{Tamemoto-Noguchi-SciRep2020,Tamemoto-Noguchi-SM2021,Tamemoto-Noguchi-PRE2022,Noguchi-SciRep2023}.

The model in Ref.\cite{Diguet-etal-PRE2024} is described by a surface model with the position variable $\vec{r}(\in\! {\bf R}^2)$ \cite{KANTOR-NELSON-PRA1987,Gompper-Kroll-PRA1992,HELFRICH-1973,Polyakov-NPB1986,Peliti-Leibler-PRL1985,Bowick-PRep2001,NELSON-SMMS2004,Wheater-JPA1994,KOIB-PRE-2005} and its moving direction $\vec{\tau}(\in\! S^1/2: {\rm half\; circle})$. This  incorporation of the membrane system $(\vec{r}, \vec{\tau})$ into the TP system $(u,v)$ entails the treatment of the space on which TP appears as a dynamical system as in the studies of Refs. \cite{Peleg-etal-PlosOne2011,Wu-etal-NatCom2018,Tamemoto-Noguchi-SciRep2020,Tamemoto-Noguchi-SM2021,Tamemoto-Noguchi-PRE2022,Noguchi-SciRep2023}. Therefore, the combination of $(u,v)$ and  $(\vec{r}, \vec{\tau})$ is considered reasonable in the sense that TPs appear on zebras and fish covered by skins, which can be considered as membranes. The reported data indicate that the TP direction is determined by the mechanical properties of the membranes, implying that the direction of the TPs can be controlled by mechanical stress or membrane stretching.

However, the surface assumed in Ref. \cite{Diguet-etal-PRE2024} is flat,  and it is necessary to extend the space ${\bf R}^2$  to ${\bf R}^3$ for $\vec{r}$ to study TPs on curved surfaces \cite{Varea-etal-PRE1999,Krause-etal-BulMatBiol2019,Krause-etal-PhiLTrRSoc2021,Vandin-etal-Softmatt2016,Nishide-Ishihara-PRL2022}. This extension is expected to present significant challenges due to the increased embedding dimension. Therefore, in this paper, we conduct a comprehensive investigation to ascertain whether the surface stretching responses, reflected in the TP direction, in the extended system remain consistent with those in the original model in Ref. \cite{Diguet-etal-PRE2024}. 

The mechanical energy supplied by the stretching process modifies the stable configuration of the unstretched membrane to another stable state. In particular, the mechanical anisotropy resulting from the stretching process gives rise to the decomposition of the membrane Hamiltonian into two distinct directional components, one parallel and the other perpendicular to the stretching direction. Therefore, the energy decomposition into the directional components is of paramount importance  in distinguishing the  {stretched states from the isotropic equilibrium state.  

This paper aims to elucidate the manner in which the directional membrane subsystem differs from the stable membrane configuration in terms of the energy localisation under the stretching and to establish a definitive relationship between the TP direction and the membrane configuration. The stretched membrane subsystem, as obtained through Monte Carlo (MC) simulation, is in a state of equilibrium. It is important to note that the stretched membrane does not exemplify a typical non-equilibrium system according to the standard criteria. Indeed, the membrane system is not open  and no energy dissipation is incorporated. 

It should be noted,  however, that the interaction strength for the internal degree of freedom (IDOF) $\vec{\tau}$ is adjustable in our modelling, and thus, real membrane configurations in the relaxation process can be simulated as equilibrium configurations  with the canonical MC simulations. In this sense, real non-equilibrium states, in the process of polymer restructuring under the stretching,  can be captured. This is one of the interesting features in Finsler geometry (FG) modelling for microscopic interactions.

\section{Triangulated Lattices \label{Sec:lattices}}
\begin{figure}[h!]
	\centering
	\includegraphics[width=8.5cm]{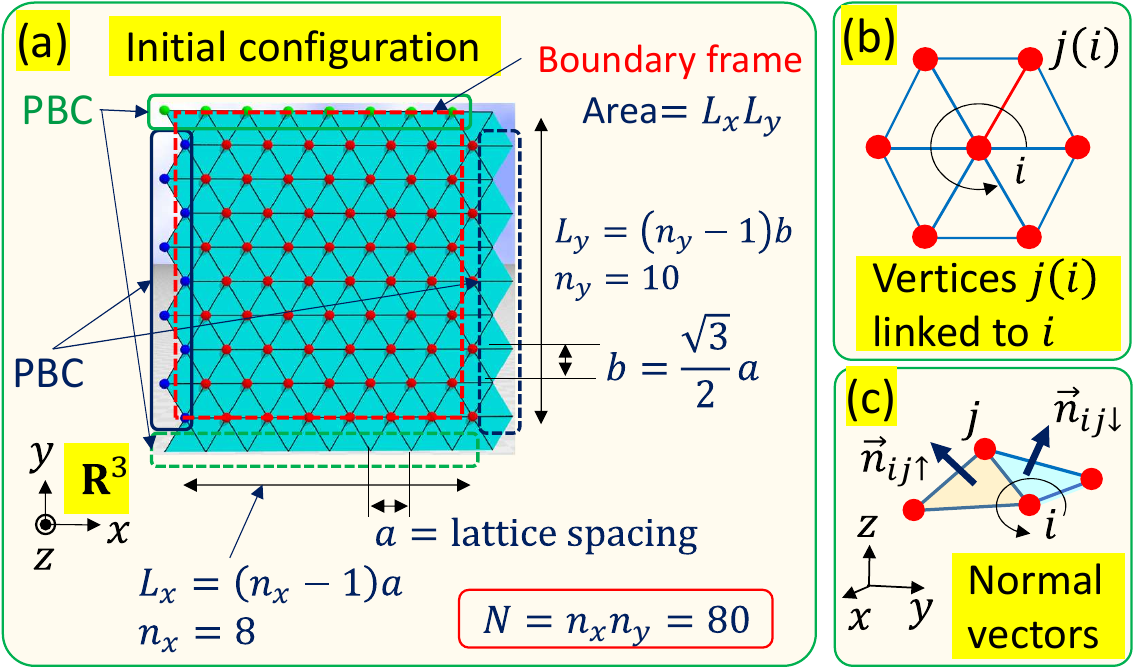}
	\caption{(a) A triangulated lattice of size $N\!=\!n_xn_y\!=\!80$ in ${\bf R}^3$, where $n_y(=\!8)$ and $n_y(=\!10)$ are relatively small to facilitate visualization of the lattice structure. In the simulations, we use the lattice of size $N\!=\!n_xn_y\!=\!50\times 58\!=\!2900$. The lattice spacing $a$ is defined as the edge length of the regular triangle. The periodic boundary condition (PBC) is assumed in both the $x$ and $y$ directions with a fixed boundary frame of side lengths $(L_x,L_y)$. (b) An illustration of vertices $j(i)$ connected to vertex $i$, and (c) normal vectors on the triangles sharing bond $ij$ (Appendix \ref{App-B}).  \label{fig-1}}
\end{figure}
We assume that the membranes are thin, and their thickness can be considered negligible. On this basis, it is reasonable to regard the membranes as two-dimensional sheets. To simulate such thin membranes, we use triangulated lattices of regular triangles for the initial configurations, imposing periodic boundary conditions (Fig. \ref{fig-1}(a)). The total number of vertices is $N\!=\!n_x n_y$, where $n_x$ and $n_y$ are the total number of vertices along $x$ and $y$ directions. To construct the lattice, we first fix $n_x$ and calculate an even number $n_y$ for the periodicity in the $y$ direction such that the edge length $L_y\!=\!(n_y-1)b$ is close to $L_x\!=\!(n_x-1)a$, where $a$ is the lattice spacing and $b\!=\!(\sqrt{3}/2)a$. As the initial configuration is composed of the regular triangles, the lattice is not completely isotropic in contrast to the randomly constructed lattices in Ref. \cite{Diguet-etal-PRE2024}. The total number of vertices $j(i)$ connected to vertex $i$ is 6 (Fig. \ref{fig-1}(b)), and the normal vectors of triangles that sharing bond $ij$ are written as $\vec{n}_{ij\downarrow}$ and  $\vec{n}_{ij\uparrow}$, which can also be written as $\vec{n}_{ji\uparrow}(=\!\vec{n}_{ij\downarrow})$ and  $\vec{n}_{ji\downarrow}(=\!\vec{n}_{ij\uparrow})$ (Fig. \ref{fig-1}(c)).  

It is possible to employ square lattices to simulate the TP system and the membrane on ${\bf R}^2$.  For curved surfaces in ${\bf R}^3$, the triangulated lattices are particularly well-suited for membrane simulations because the triangles remain peacewise linear for any surface deformations. This linear property is useful for the calculations of geometric quantities such as the area and curvature. Consequently, we utilise the triangulated lattice in this paper \cite{note-2}.

 \section{Finsler geometry modelling of anisotropic diffusion \label{Sec:FG-model}}
In this section, we introduce a new IDOF $\vec{\tau}$, which is different from the vertex-fluctuation direction introduced in Ref. \cite{Diguet-etal-PRE2024}, to define anisotropic diffusion in RD equation for TP and anisotropic mechanical strength for tensile and bending deformations in membrane Hamiltonian.

\subsection{Reaction diffusion equations \label{Sec:RD-equation}}
Let $u$ and $v$ be the activator and inhibitor, respectively, and satisfy the following FitzHugh-Nagumo type reaction-diffusion (RD) equation:
\begin{eqnarray}
	\begin{split}
		\label{FN-eq-Eucl}
		&\frac {\partial u}{\partial t}=D_u \Laplace({\tau})u +f(u ,v), \quad  f=u -u ^3-v, \\
		&\frac {\partial v}{\partial t}=D_v  \Laplace({\tau}) v+ g(u,v), \quad g= \gamma(u -\alpha v),
	\end{split}
\end{eqnarray}
where the diffusion coefficients $D_u, D_v$  and the constants $\alpha (>\!0), \gamma (>\!0)$ are suitably fixed. The Laplacian $\Laplace({\tau})$ depends on IDOF $\vec{\tau}$ describing anisotropic diffusion for TPs in FG modelling \cite{Koibuchi-Sekino-PhysA2014,Proutorov-etal-JPC2018,Diguet-etal-CMS2024,SS-Chern-AMS1996}.  The discrete RD equation is provided below, and the anisotropic diffusion $\Laplace({\tau})$ is solved numerically under the condition that the IDOF $\vec{\tau}$ is suitably given as a membrane configuration produced by MC simulation. 

\subsection{Internal degree of freedom for directions of polymer}
\begin{figure}[h!]
	\centering
	\includegraphics[width=8.5cm]{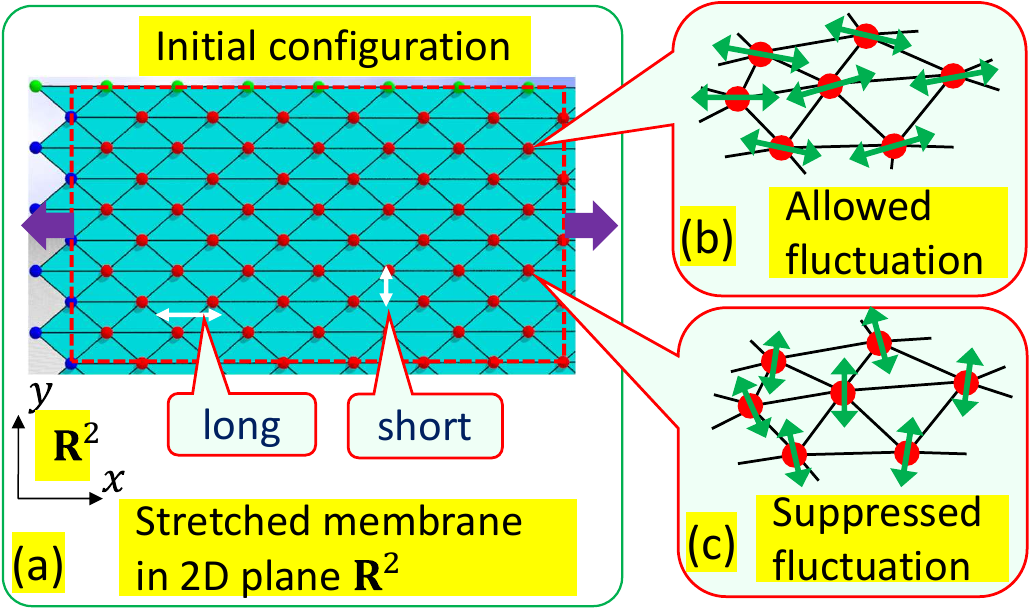} 
	\caption{(a) A stretched lattice  along the $x$-axis in 2D plane ${\bf R}^2$. The bond length parallel  to the $x$-axis is expected to be longer than that of the perpendicular bonds. Consequently, the distance between vertices along the $y$-axis is relatively shorter compared with that along the $x$-axis, resulting in Helfrich repulsion along the $y$-axis. This repulsion (b) allows for the vertex fluctuation  in the $x$ direction and (c) suppresses it in the $y$ direction.
	\label{fig-2}}
\end{figure}
Anisotropic TPs are closely related to anisotropic diffusion \cite{Kondo-Nature1995,Shoji-etal-DevDyn2003,Iwamoto-Shoji-RIMS2018}. It was reported in \cite{Diguet-etal-PRE2024} that the anisotropic diffusion is dynamically implemented in Laplace operators by FG modelling technique and the pattern direction is controlled by a tensile deformation of membranes (Fig. \ref{fig-2}(a)). This controllability  is due to the alignment of the IDOF $\vec{\tau}$, which is the vertex fluctuation direction in \cite{Diguet-etal-PRE2024}, along the tensile deformation direction (Fig. \ref{fig-2}(b)). The orientation of the IDOF $\vec{\tau}$ is as a result of the suppression of thermal fluctuations along the direction perpendicular to the stretching direction (Fig. \ref{fig-2}(c)).  This suppression is due to the Helfrich repulsion \cite{Helfrich-NatF1978,Evans-Parsegian-PNAS1986}, which originates from the excluded volume effect expected for the self-avoiding membranes in the  two-dimensional plane ${\bf R}^2$ (Fig. \ref{fig-3}(a)) \cite{Chen-etal-SM2015}.  
\begin{figure}[h!]
	\centering
	\includegraphics[width=8.5cm]{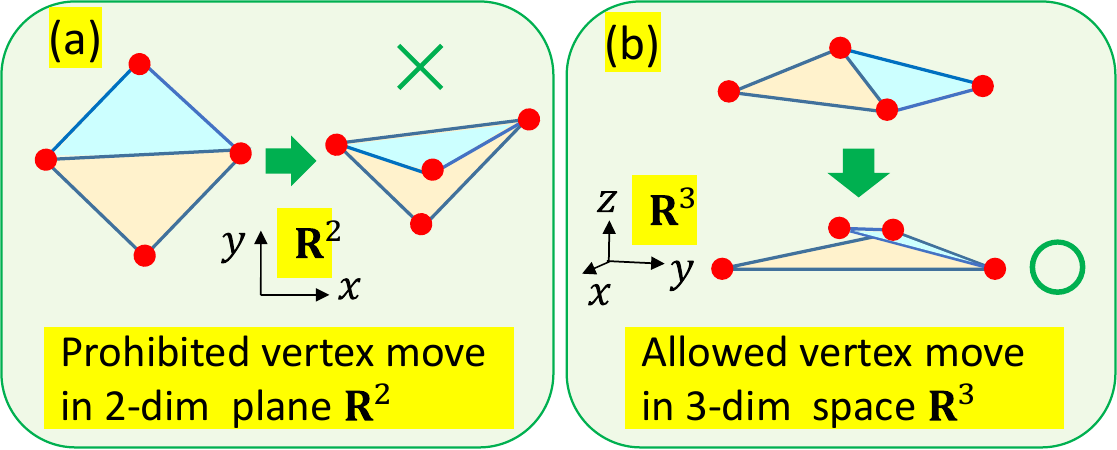}
	\caption{(a) A forbidden vertex move ($\times$) in self-avoiding surfaces in ${\bf R}^2$. (b) A configuration that is forbidden in ${\bf R}^2$ is permitted ($\bigcirc$) in ${\bf R}^3$ due to the out-of-plane surface deformation.  
	\label{fig-3}}
\end{figure}

However, for membranes in ${\bf R}^3$, the surface self-avoidance in ${\bf R}^2$  will no longer be a factor to the orientation of $\vec{\tau}$  due to a new dimension. As illustrated in  Fig. \ref{fig-3}(b), the $z$-axis for vertex motion effectively eliminates the surface self-avoidance, and therefore no repulsion is expected in the direction perpendicular to the stretching direction, even though the surface is almost flat due to the boundary frame in ${\bf R}^3$.  Accordingly, the IDOF $\vec{\tau}$ defined by the fluctuation direction for FG modelling of membranes in Ref. \cite{Diguet-etal-PRE2024} is not applicable in ${\bf R}^3$. 
\begin{figure}[h!]
	\centering
	\includegraphics[width=8.5cm]{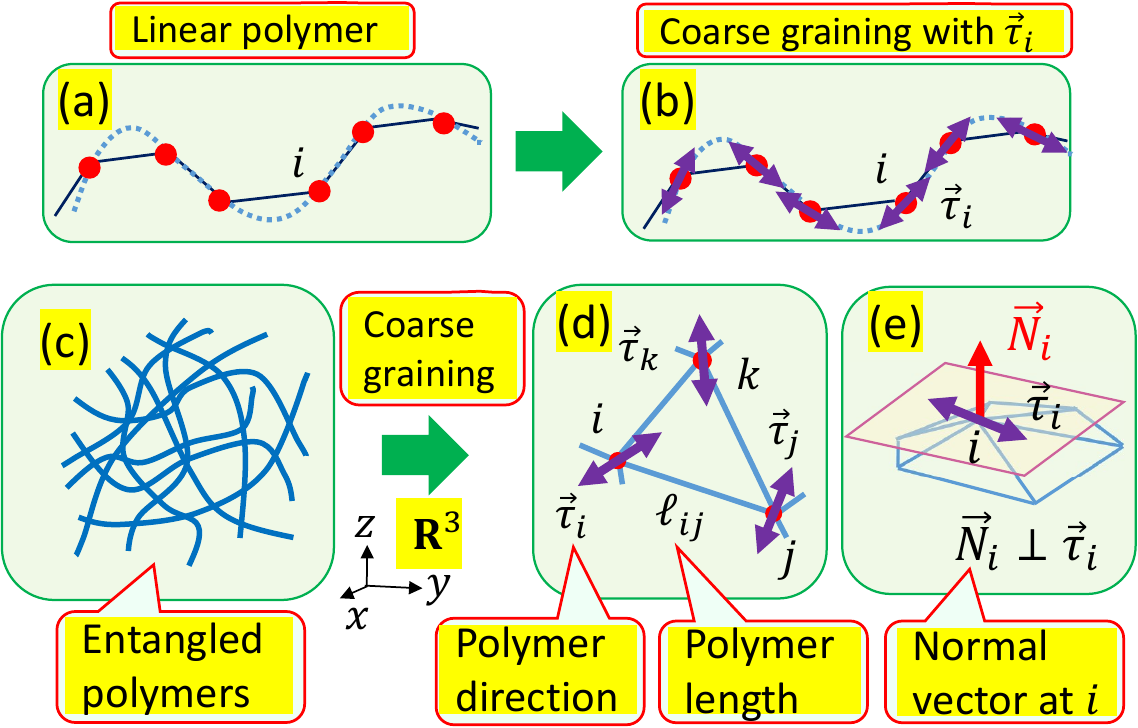}
	\caption{Illustrations of (a) a linear polymer, and (b) a coarse-grained linear polymer with the polymer direction $\vec{\tau}$. (c) An illustration of the entanglement of polymers in polymerized membranes. For a coarse-grained modelling of polymerized membranes, we assume two independent variables: (d) the polymer direction $\vec{\tau}_i (\in\!S^2/2)$ (half-sphere) at vertex $i$ and the polymer length $\ell_{ij}(=\!\|\vec{r}_i\!-\!\vec{r}_j\|)$  of bond $ij$ connecting vertex positions $\vec{r}_i, \vec{r}_j  (\in\! {\bf R}^3)$. (e) $\vec{\tau}_i (\in\!S^2/2)$ at vertex $i$  is defined such that $\vec{\tau}_i\!\perp\! \vec{N}_i$, where $\vec{N}_i$ is the normal vector of a tangent plane at vertex $i$. 
	\label{fig-4}   }
\end{figure}

To find a new IDOF, we illustrate a chain model for linear polymers and its coarse-grained model by introducing a polymer direction  $\vec{\tau}_i$ in Figs. \ref{fig-4}(a),(b) \cite{MDoi-Edwards-1986,MDoi-2013}. 
Analogously, we extend the triangulated surface model to a coarse-grained lattice model of entangled polymers  (Fig. \ref{fig-4}(c)) by employing the polymer direction $\vec{\tau}_i$ in conjunction with the polymer length $\ell_{ij}$ (Fig. \ref{fig-4}(d)). In the lattice modelling, the polymer length $\ell_{ij}$ is defined as the distance between the nearest neighbour vertices of positions $\vec{r}_i(\in\!{\bf R}^3)$ and $\vec{r}_j$ such that $\ell_{ij}\! =\!\|\vec{r}_i\!-\!\vec{r}_j\|$ \cite{KANTOR-NELSON-PRA1987,Gompper-Kroll-PRA1992,HELFRICH-1973,Polyakov-NPB1986,Peliti-Leibler-PRL1985,Bowick-PRep2001,NELSON-SMMS2004,Wheater-JPA1994,KOIB-PRE-2005}. We define the polymer direction $\vec{\tau}_i (\in S^2/2)$ on a tangential plane at vertex $i$ (Fig. \ref{fig-4}(e)). The tangential plane is given by the normal vector $\vec{N}_i$, which is defined by 
\begin{eqnarray}
	\label{normal-vector}
	\vec{N}_i\!=\!\sum_{j(i)} A_{j(i)}\vec{n}_{j(i)}/\sum_{j(i)}A_{j(i)},
\end{eqnarray}
where $A_{j(i)}$ is the area of the triangle $j(i)$ connected to vertex $i$, and $\vec{n}_{j(i)}$ is the normal vector.

\subsection{Hamiltonians for membranes with Turing patterns \label{sec:total-discrete-H}}
The variable $\vec{\tau}(\in S^2/2)$, which has only in-plane components of the tangential plane at the vertices, is updated by MC technique using the Hamiltonian
\begin{eqnarray}
	\label{discrete-total-Hamiltonian}
		H(\vec{r},\vec{\tau})=H_1+\kappa H_2+U_V+\lambda H_\tau 
\end{eqnarray}
instead of using $H(\vec{r},\vec{\tau},u,v)=H_1+\kappa H_2+U_V+\lambda H_\tau+\left(H_{u}-\frac{1}{\gamma}H_{v}\right)$. We assume $k_BT\!=\!1$ for the unit of energy, where $k_B$ and $T$ are the Boltzmann constant and the temperature. This energy condition encompasses both cases of $k_B\!=\!1$ and $T\!=\!1$, which are assumed for the evaluation of thermodynamic entropy below.  The energy scale of $H_{u}-\frac{1}{\gamma}H_{v}$ is presumed to be equivalent to that of membrane Hamiltonian. However, the term $H_{u}\!-\!\frac{1}{\gamma}H_{v}$ is not included in $H(\vec{r},\vec{\tau})$ of Eq. (\ref{discrete-total-Hamiltonian}) due to the negligible magnitude of $H_{u}$ and $(1/\gamma)H_{v}$, which are approximately $10^{-2}$ to $10^{-3}$ smaller than the other terms.  The rationale for considering  the potential inclusion of $H_{u}\!-\!\frac{1}{\gamma}H_{v}$ is that  this term depends on $\vec{\tau}$, and the steady-state equations in Eq. (\ref{FN-eq-Eucl}) are derived from  $H_{u}-\frac{1}{\gamma}H_{v}$   (Appendix \ref{App-A}). 

The length scale of the TP system under investigation is defined by the size of the periodic pattern, which can be macroscopic as well as the membranes composed of thermally fluctuating polymers. The scale of the TP should be sufficiently larger than that of the polymer. The lattice spacing $a$ corresponds to the distance between polymer positions in a coarse-grained sense, and therefore, the constraint on the size $a$ is that $a$ is larger than the monomer size (typically  $\sim 10^{-9} {\rm m}$). 

It should be noted that  $H_{u}\!-\!\frac{1}{\gamma}H_{v}$ is negligibly small compared with the other terms in $H$  of Eq. (\ref{discrete-total-Hamiltonian}), as mentioned above. For this reason, the numerical results are independent of whether the term $H_{u}\!-\!\frac{1}{\gamma}H_{v}$ is included or not in $H$ for the MC updates of $\vec{r}$ and $\vec{\tau}$. This indicates that the membrane configurations, represented by $(\vec{r}, \vec{\tau})$, are not influenced by $(u,v)$, in sharp contrast to the fact that the TP directions are determined by the membrane stretching. This point is in contrast to the studies in Refs. \cite{Tamemoto-Noguchi-SciRep2020,Tamemoto-Noguchi-SM2021,Tamemoto-Noguchi-PRE2022,Noguchi-SciRep2023}, where the variable $u$ is assumed to have a direct coupling with membranes in their models.  
It is also important to note that the constraining potential for fixing the membrane with the boundary frame is not included in the Hamiltonian; the TP system on membranes is not completely isolated because the membranes are fixed to the external space with the boundary frame.

The expressions of the terms on the right hand side of  $H$ in Eq. (\ref{discrete-total-Hamiltonian}) are as follows:
	\begin{eqnarray}
	\label{discrete-Hamiltonian}
	\begin{split}
		&H_1=\sum_{ij}\Gamma_{ij}^G(\tau)\ell_{ij}^2,\quad \ell_{ij}^2\!=\!\|\vec{r}_i\!-\!\vec{r}_j\|^2, \\ 
		&H_2=\sum_{ij}\Gamma_{ij}^b(\tau)(1-\vec{n}_{ij\downarrow}\cdot \vec{n}_{ij\uparrow}),\\
		&U_V=\sum_{ij}U_{ij}, \quad U_{ij}=\left\{ \begin{array}{@{\,}ll}0\quad  (\ell_{\min}\leq \ell_{ij}\leq \ell_{\rm max}) \\
			\infty\quad  ({\rm otherwise})
		\end{array} 
		\right., \\
		&\quad(\ell_{\min}=0.01a  \quad\ell_{\rm max}=3a,\quad a : {\rm latt.\; sp.}),\\
		&  H_\tau=\frac{3}{2}\sum_{ij}\left(1-\left(\vec{\tau}_i\cdot \vec{\tau}_j\right)^2\right),
	\end{split}
\end{eqnarray}
The first term $H_1$ is the Gaussian bond potential defined by the sum of the bond length squares $\ell_{ij}^2$ with $\Gamma_{ij}^G(\tau)$ which depends on the polymer directions $\vec{\tau}$ \cite{note-3}. Note that $\sum_{ij}$ denotes the sum of bonds $ij$ and is the same as $\sum_{i>j}$ for the sum over vertices $i$ and $j$ connected by a bond.  The second term $\kappa H_2$ is the bending energy with the bending rigidity $\kappa$, where $\vec{n}_{ij\downarrow}, \vec{n}_{ij\uparrow}$ denote unit normal vectors of triangles sharing bond $ij$ (Fig. \ref{fig-1}(c)). The third term $U_V$ is a constraint potential for the bond length with the maximum and minimum lengths $\ell_{\rm max}$ and $\ell_{\rm min}$ defined by the lattice spacing $a$  (Fig. \ref{fig-1}(a)). The fourth term $H_\tau$ is the self-interaction or correlation energy for the nearest neighbour pairs of  $\vec{\tau}$ connected by a bond \cite{note-4}. The interaction between the variables $\vec{\tau}$ is significantly influenced by $H_\tau$ for a large coefficient $\lambda$. Note that  $H_\tau$ is symmetric under the rotation of $\vec{\tau}$, and therefore, the $\vec{\tau}$ direction is not directly influenced by $\lambda H_\tau$. Rather, the direction of $\vec{\tau}$ is primarily influenced by the interactions between the intensive and extensive components, such as $\Gamma_{ij}^G(\tau)$ and  $\ell_{ij}^2$ etc., in $H_1$ and $H_2$.

The terms $H_u$ and $H_v$ in $H=H_u\!-\!(1/\gamma)H_v$ correspond to the right-hand sides of Eq. (\ref{FN-eq-Eucl}) and are composed of the diffusion terms $H_{u,v}^D$ and the reaction terms $H_{u,v}^R$ (Appendix \ref{App-C})
 	\begin{eqnarray}
 	\label{discrete-Hamiltonian-for-TP}
 	\begin{split}
 		&H_{u}=D_uH_u^D+H_u^R, \quad H_u^D= \sum_{ij}D_{ij}^u(\tau)\left(u _j-u _i\right)^2, \\
 		&H_{v}=D_vH_v^D+H_v^R, \quad H_v^D=\sum_{ij}D_{ij}^v(\tau)\left(v _j-v _i\right)^2. 
 	\end{split}
 \end{eqnarray}
In this context,  the symbols $D_{ij}^u$ and $D_{ij}^v$ are employed to represent the effective and microscopic diffusion coefficients in $H_{u,v}^D$. It is presumed that there is no risk of confusion with the constants $D_u$ and $D_v$.

Detailed information on the effective coupling constants $\Gamma_{ij}^{G,b}(\tau)$ and $D_{ij}^{u,v}(\tau)$ is given in Appendix \ref{App-B}. Here we describe the unit Finsler length $\chi_{ij}$ of bond $ij$ from vertices $i$ to $j$ (Fig.\ref{fig-4}(d)) assumed in $\Gamma_{ij}^{G,b}(\tau)$ and $D_{ij}^{u,v}(\tau)$:
\begin{eqnarray}
	\label{Finsler-unit-lengths-1}
	\begin{split}
		&\left\{ \begin{array}{@{\,}ll}
			\chi_{ij}^G=\sqrt{1-|\vec{\tau}_i\cdot\vec{e}_{ij}|^2} + \chi_0,	\; ({\rm sin}) \\
			\chi_{ij}^b=|\vec{\tau}_i\cdot\vec{e}_{ij}| + \chi_0, \; ({\rm cos}) \\
			\chi_{ij}^u=|\vec{\tau}_i\cdot\vec{e}_{ij}| + \chi_0, \; ({\rm cos}) \\
			\chi_{ij}^v=\sqrt{1-|\vec{\tau}_i\cdot\vec{e}_{ij}|^2} + \chi_0,\; ({\rm sin})
		\end{array} \right., 	\quad ({\rm model\; 1})
	\end{split}
\end{eqnarray}
\begin{eqnarray}
	\label{Finsler-unit-lengths-2}
	\begin{split}
		&\left\{ \begin{array}{@{\,}ll}
			\chi_{ij}^G=|\vec{\tau}_i\cdot\vec{e}_{ij}| + \chi_0, \; ({\rm cos}) \\
			\chi_{ij}^b=\sqrt{1-|\vec{\tau}_i\cdot\vec{e}_{ij}|^2} + \chi_0,	\; ({\rm sin}) \\
			\chi_{ij}^u=|\vec{\tau}_i\cdot\vec{e}_{ij}| + \chi_0, \; ({\rm cos}) \\
			\chi_{ij}^v=\sqrt{1-|\vec{\tau}_i\cdot\vec{e}_{ij}|^2} + \chi_0,\; ({\rm sin})
		\end{array} \right.,  \quad ({\rm model\; 2})
	\end{split}
\end{eqnarray}
where $\vec{e}_{ij}(=\!(\vec{r}_j\!-\!\vec{r}_i)/\|\vec{r}_j\!-\!\vec{r}_i\|)$ denotes the unit tangent vector from vertices $i$ to $j$. The $\chi_0$ is a small number,  which can control the strength of anisotropy interaction, and will be discussed in the following section. The $\chi_{ij}^{G,b,u,v}$ are of two types; sine and cosine, denoted by the symbols (sin) and (cos). We define two variations in the modelling; model 1 and model 2, where the variables $\chi_{ij}^G$ and $\chi_{ij}^b$ are fixed in two different combinations $\chi_{ij}^G({\rm sin}), \chi_{ij}^b({\rm cos})$ for model 1 and $\chi_{ij}^G({\rm cos}), \chi_{ij}^b({\rm sin})$ for model 2, while $\chi_{ij}^u$ and $\chi_{ij}^v$ are fixed to cosine and sine types, respectively, for both models. In the model of Ref. \cite{Diguet-etal-PRE2024}, the FG modelling is only applied  to the diffusion terms $H_u^D$ and $H_v^D$, and there is no variation as in models 1 and 2. 

A large difference in $D_u$ and $D_v$ is necessary for anisotropic TPs \cite{Kondo-Nature1995,Shoji-etal-DevDyn2003,Iwamoto-Shoji-RIMS2018}. In addition to this difference, directional diffusions of $u$ and $v$ are also necessary\cite{Shoji-etal-DevDyn2003,Iwamoto-Shoji-RIMS2018}, and such anisotropies are dynamically generated and reflected in $D_{ij}^{u,v}(\tau)$ depending on the $\vec{\tau}$ direction, which is controlled by a membrane stretching \cite{Diguet-etal-PRE2024}. These mechanical anisotropies implemented in $\Gamma_{ij}^{G}(\tau)$ and $\Gamma_{ij}^{b}(\tau)$ lead to the notions of "easy axes for tensile and bending deformations", along which anisotropic TPs appear. This is one of the main results in this paper.

\begin{figure}[h!]
	\centering
	\includegraphics[width=8.5cm]{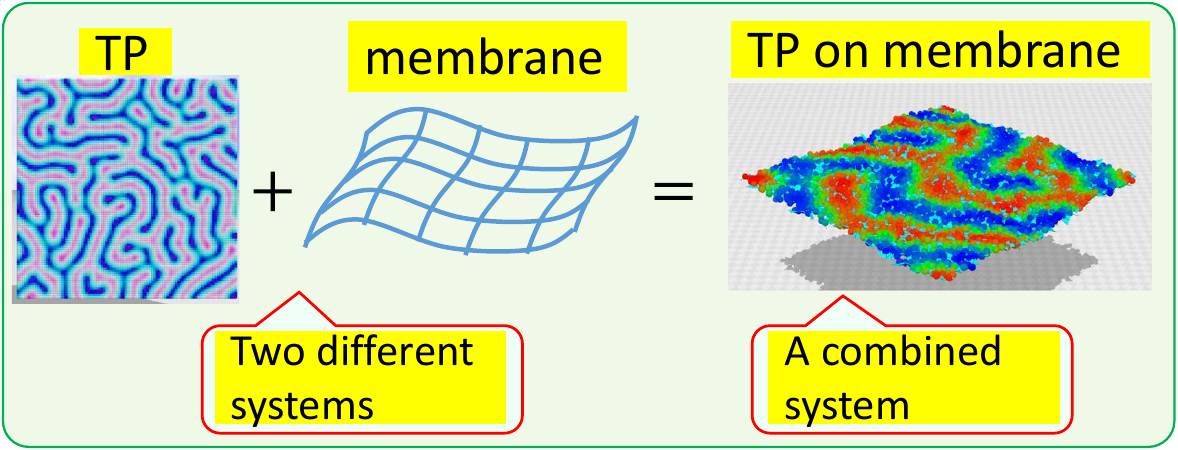}
	\caption{TP and membranes, which are originally two different systems, are coupled with each other to form a combined system, designated as  "TP on membranes". This combined system exhibits one-way interaction, whereby the TP direction is determined by the mechanical properties of membranes. 
		\label{fig-5}}
\end{figure}
Finally, in this subsection we illustrate the model of TPs on membranes in Fig. \ref{fig-5}; two different dynamical systems are contained. The model of TPs described by the RD equation for $(u,v)$ in Eq. (\ref{FN-eq-Eucl}) and the membranes described by the Hamiltonian $H$ in Eq. (\ref{discrete-total-Hamiltonian}) are originally independent, and these two models couple with each other to form a combined system such as models 1 and 2. Therefore, we expect that (i) the TP system is influenced by the membranes such that the TP direction is determined by the mechanical properties of the membranes, and conversely (ii) the mechanical properties of the membranes are influenced by the TP directions. However, as mentioned above, (ii) is not always true.

\subsection{Formula for surface tension and the thermodynamic entropy of stretched membranes}
\begin{figure}[h!]
	\centering
	\includegraphics[width=7.5cm]{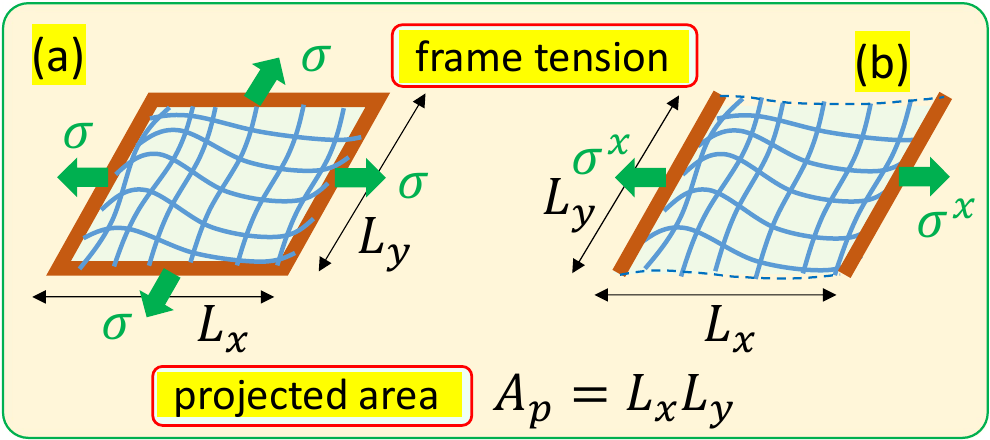}
\caption{ 
(a) Frame tension $\sigma$ of membranes extending across the frame of square boundary. (b) Directional frame tension $\sigma^x$ of membranes extending across the  parallel boundaries, with the movable frame in view. Note that the projected area of the frame is given by $A_p$ in both (a) and (b).  The term "surface tension" is employed for simplicity.
		\label{fig-6}}
\end{figure}
In this subsection, the directional surface tension is introduced using the notion of  direction-dependent energy (see Appendix \ref{App-D}). Comprehensive details pertaining to the isotropic surface tension $\sigma$ and the directional surface tensions $\sigma^\mu, (\mu\!=\!x,y)$  can be found in Appendices \ref{App-E} and \ref{App-F} respectively.

The surface tension $\sigma$ is microscopically evaluated by assuming the membrane as a surface of fixed area $A_p$  without the internal structure instead of using the unknown potential for the fixed boundary in the microscopic partition function  \cite{Wheater-JPA1994} (Appendix \ref{App-E}). Undulating membranes are assumed to extend over the square frame of area $A_p$, which is smaller than the membrane area, and hence, $\sigma$ is the frame tension (Fig. \ref{fig-6}). We call this $\sigma$  the surface tension in this paper. The formula for $\sigma$ of membranes in ${\bf R}^{2}$ spanning the boundary frame of the projected area $A_p$ is given by
\begin{eqnarray}
	\label{surface-tension}
	\begin{split}
		&\sigma=\frac{N}{A_p}\left(\frac{1}{N}\sum_{ij}\Gamma_{ij}^G\ell_{ij}^2-\left(1-\frac{N_{\rm fix}}{2N}\right)\right), \quad ({\rm for \; models\; in\;}{\bf R}^{2}),\\ 
		&A_p=L_xL_y=(n_x-1)(n_y-1)\frac{\sqrt{3}a^2}{2}, \; a=0.525,
	\end{split}
\end{eqnarray}
where $N_{\rm fix}\!=\!n_x\!+\!n_y$, and the lattice spacing $a(=\!0.525)$ is the same as in Ref. \cite{Diguet-etal-PRE2024}.

It is reasonable to define the directional surface tension $\sigma^\mu, (\mu=x,y)$ using the directional energy decomposition of $H_1$ as follows (Appendix \ref{App-F}). 
\begin{eqnarray}
	\label{direc-dep-surface-tension-2D}
	\begin{split}
		\sigma^\mu=&\frac{2N}{A_p}\left(\frac{1}{N}\Gamma_\mu^GH_1^\mu-\frac{N-N_{\rm fix}+n_\mu}{2N}\right), \; (\mu=x,y), \\
		&({\rm for \; models\; in\;}{\bf R}^2).
	\end{split}
\end{eqnarray}

In the case of the models in ${\bf R}^3$, we calculate the surface tension by
\begin{eqnarray}
	\label{direc-dep-surface-tension-3D}
	\begin{split}
		\sigma^\mu=&\frac{2N}{A_p}\left(\frac{1}{N}\Gamma_\mu^GH_1^\mu+\frac{1}{2N}\Delta H_2-\frac{N-N_{\rm fix}+n_\mu}{2N}\right), \; (\mu=x,y), \\
		& \Delta H_2=\kappa \left.\frac{\partial H_2(\alpha x,y,z)}{\partial \alpha} \right|_{\alpha=1}, \quad ({\rm for \; models\; in\;}{\bf R}^3).
	\end{split}
\end{eqnarray}

Using the free energy $F$ in Eq. (\ref{free-energy-F-dir}) of membranes in Appendix \ref{App-F}, we obtain the formula for the entropy $S$ such that
\begin{eqnarray}
	\label{entropy}
		S(R_{xy})=(1/T)(U-F)=U-F,
\end{eqnarray}
where the temperature $T$ is fixed to $T\!=\!1$. Let us assume that the surface area is expanded from $A_p$ to $A_P\!+\!\delta A_p$ by moving the frame along the $x$ direction (Fig \ref{fig-6}(b)). Then, the free energy variation is given by $\delta F\!=\!\frac{\partial F}{\partial A_p}\delta A_p\!=\!\frac{\partial}{\partial A_p}\left(\int^{A_p}\sigma^x(A)dA\right)\delta A_p\!=\!\sigma^x(A_p)\delta A_p$. The variation of internal energy $\delta U$ can also be expressed as $\delta U\!=\!\frac{\partial U}{\partial A_p}\delta A_p\!=\!\frac{\partial}{\partial A_p}\left(\int^{A_p}u(A)dA\right)\delta A_p\!=\!u(A_p)\delta A_p$ using the internal energy density $u$ of the continuous surface. Here, we replace the density $u(A_p)$ with the membrane energy per unit area, and we have $ u(A_p)\!=\!(H_1 \!+\!\lambda H_\tau)/A_p$. Note that the lower values in the integral for $S$, $U$ and $F$ are not explicitly stated, since  the focus is on the variations rather than the quantities themselves. It is clear that $A_p\sigma^x\!=\!2\Gamma_x^GH_1^x\!-\!(N\!-\!n_y)$ from Eq. (\ref{direc-dep-surface-tension-2D}), and therefore, we have  $\delta S(R_{xy})\!=\!\left((H_1\!+\!\lambda H_\tau)/A_p-\!(2\Gamma_x^GH_1^x\!-\!(N\!-\!n_y))/A_p\right)\delta A_p$. Thus,
\begin{eqnarray}
	\label{entropy-2-dim}
\begin{split}
	&\delta S(R_{xy})=s(A_p)\delta A_p, \quad ({\rm for \; models\; in\;}{\bf R}^2),\\
	&s(A_p)=\frac{H_1+\lambda H_\tau-2\Gamma_x^GH_1^x+N-n_y}{A_p}.
\end{split}
\end{eqnarray}

	In the case of the models in ${\bf R}^{3}$,  
	the expression of $\delta S(R_{xy})$ using $\sigma^x$ in Eq. (\ref{direc-dep-surface-tension-3D}) 
	is given by
	\begin{eqnarray}
		\label{entropy-3-dim}
		\begin{split}
			&\delta S(R_{xy})=s(A_p)\delta A_p, \quad ({\rm for \; models\; in\;}{\bf R}^3),\\
			&s(A_p)=\frac{H_1+\kappa H_2+\lambda H_\tau-2\Gamma_x^GH_1^x-\Delta H_2+N-n_y}{A_p}, 
		\end{split}
	\end{eqnarray}
where $\kappa H_2$ is included in the internal energy, and $\Delta H_2$ is given in Eq. (\ref{direc-dep-surface-tension-3D}).

Thus, we find that the entropy variation $\delta S$ with respect to $\delta A_p$ is proportional to the entropy density $s(A_p)$.  We can calculate this  $s(A_p)$ by varying $\lambda$ and $R_{xy}$. Under the variation of $R_{xy}$, the area  $A_p$ is assumed to remain constant \cite{note-5}. Therefore, $s(A_p)$ can also be expressed as $s(R_{xy})$.

It is important to acknowledge the dependence of $s(R_{xy})$ on $\lambda$. Variations in $\lambda$ result in  alterations to the $\vec{\tau}$ configuration and, consequently, to the anisotropic mechanical properties of membranes (see the supplementary material (1) for further insights).

Additionally, if $s(R_{xy})$ is significant then the entropy change under $\delta A_p$ from the current $S(A_p)$ becomes significant. Therefore, information regarding $s(R_{xy})$ as a function of $\lambda$ is interesting. Specifically, the maximal entropy condition for the $\lambda$ variation such that $\frac{\partial S(A_p;\lambda)}{\partial \lambda}\!=\!0$ at $\lambda\!=\!\lambda_c$ is satisfied if the following condition for the entropy density is satisfied
\begin{eqnarray}
\label{entropy-convex-criteria}
		 \frac{\partial s}{\partial \lambda}(R_{xy};\lambda_c)=0\left(\Leftrightarrow s(R_{xy};\lambda)\;\; {\rm has\;\; a \;\; peak\;\; point\;\;}\lambda_c\right)
\end{eqnarray}
within a specified range of $\lambda$. 
Indeed, since the entropy is expressed in the integral form as   $S(A_p;\lambda)\!=\!\int^{A_p}s(A;\lambda)dA$, it is straightforward to find that $\frac{\partial}{\partial \lambda}S(A_p;\lambda_c)\!=\!\frac{\partial}{\partial \lambda}\int^{A_p}s(A;\lambda)dA|_{\lambda=\lambda_c}\!=\!\int^{A_p}\frac{\partial s}{\partial \lambda}(A;\lambda_c)dA$. 
Henceforth, we will refer to the entropy density $s(A_p)\!=\!\delta S(R_{xy})/\delta A_p$ as "entropy".

\section{Numerical technique \label{Sec:numerical-tech}}
\subsection{Discrete RD equation and Monte Carlo}
Steady-state solutions of the RD equation for $u$ and $v$ are obtained by the following time evolution of first order with respect to the discrete time ${\Delta} t$
\begin{eqnarray}
	\label{discrete-t-iterations-tlattice}
	\begin{split}
		&u_{i}(t+{\Delta} t)\leftarrow u_{i}(t) +{\Delta} t \left[D_u{\Laplace}({\tau})u_{i}(t) +f(u_{i}(t),v_{i}(t)) \right],\\
		&v_{i}(t+{\Delta} t)\leftarrow v_{i}(t) +{\Delta} t \left[D_v{\Laplace}({\tau})v_{i}(t) +g(u_{i}(t),v_{i}(t))\right], 
	\end{split}
\end{eqnarray}
which correspond to the RD equation in Eq. (\ref{FN-eq-Eucl}). The discrete RD equation with Finsler metric is given in Eq. (\ref{discrete-RD-Fins}) in Appendix \ref{App-C}, where the Laplacians are given by 
\begin{eqnarray}
	\label{discrete-Laplace}
	\begin{split}
	&\Laplace({\tau}) u_i = 2\sum_{j(i)}D_{ij}^u(\tau)\left(u_j-u_i \right), \\
	&\Laplace({\tau}) v_i = 2\sum_{j(i)}D_{ij}^v(\tau)\left(v_j-v_i \right).
	\end{split}
\end{eqnarray}

The vertex position $\vec{r}_i(\in\!{\bf R}^3)$ and IDOF $\vec{\tau}$ are updated by the Metropolis MC technique using the Hamiltonian $H$ in Eq. (\ref{discrete-total-Hamiltonian}). A new position $\vec{r}_i{^\prime}\!=\!\vec{r}_i\!+\!\Delta\vec{r}$ is accepted with the probability ${\rm Min}[1, \exp(-\Delta H)]$, where $\Delta H\!=\!H(\vec{r}_i{^\prime})\!-\!H(\vec{r}_i)$. The length of the small random vector $\Delta\vec{r}$ is set appropriately for a reasonable value of the acceptance rate. A new value $\vec{\tau}{^\prime}$, which is independent of the current one $\vec{\tau}$, is accepted with the probability ${\rm Min}[1, \exp(-\Delta H)]$ with $\Delta H\!=\!H(\vec{\tau}{^\prime})\!-\!H(\vec{\tau})$. 

Here, we describe the stretching ratio $R_{xy}$ for fixing the boundary frame. As shown in Fig. \ref{fig-1}(a), the lattice size is characterized by $L_x$ and $L_y$. Using these $L_x$ and $L_y$, a lattice deformation is defined by $R_{xy}$ such that 
\begin{eqnarray}
	\label{lattice-deformation}
	L_x\to L_x^\prime=L_x/\sqrt{R_{xy}},\quad L_y\to L_y^\prime=\sqrt{R_{xy}}L_x,
\end{eqnarray}
where $L_x^\prime L_y^\prime=L_x L_y$ is satisfied.  Therefore, it is crucial to note that the frame area remains constant under this lattice deformation. Note  also that $R_{xy}\!=\!(L_y^\prime/L_x^\prime)/(L_y/L_x)$, which is  not exactly the same as $L_y^\prime/L_x^\prime$, because $L_y/L_x\!\not=\!1$  in general (Fig. \ref{fig-1}(a)).

\subsection{Hybrid numerical technique \label{hybrid-num}}
We have four variables; $\vec{r}_i$ and $\vec{\tau}_i$ for membranes, and $u_i$ and $v_i$ for TPs. The membrane configurations described by $\vec{r}_i$ and $\vec{\tau}_i$ are updated by Metropolis Monte Carlo technique, and the TPs described by $u_i$ and $v_i$ are updated by discrete RD equations. These techniques are combined to obtain steady-state configurations. The variables $u_i, v_i$ and $\vec{r}_i$, $\vec{\tau}_i$ are updated as follows:
\begin{enumerate}
	\item [(i)]  The variables $u_i(t)$ and $v_i(t)$ are incremented to $u_i(t+{\Delta}t)$ and $v_i(t+{\Delta}t)$ for $i=1,\cdots,N$ by Eq. (\ref{discrete-t-iterations-tlattice}) with $\Delta t\!=\!0.001$. Initial configurations $(u_i(0),v_i(0))$ are randomly fixed near zero $(0,0)$.
	\item [(ii)] MC updates for $\vec{r}_i$ are performed for $i=1,\cdots,N$ once for each $i$, and then MC updates for $\vec{\tau}_i$ are performed for $i=1,\cdots,N$ for the same manner. The initial configuration of $\vec{r}_i$ is fixed to that of the initial lattice configuration (Fig. \ref{fig-2}(a)), and the initial  $\vec{\tau}_i$ is randomly fixed. 
	\item [(iii)] Steps (i) and (ii) are consecutively repeated $n_{\rm MC}$ times. $n_{\rm MC}\!=\!2\!\times\!10^5$ (models in ${\bf R}^2$) and $n_{\rm MC}\!=\!3\!\times\!10^5$ (models in ${\bf R}^3$) for the $N\!=\!2900$ lattice with $n_x\!=\!50$, $n_y\!=\!58$ (Fig. \ref{fig-1}(a)).
	\item [(iv)] Step (i) is repeated with the final configurations of $\vec{r}_i$ and $\vec{\tau}_i$ obtained in step (iii) until the following conditions are satisfied:
	\begin{eqnarray}
		\begin{split}
			\label{convergence-FG}
			&{\rm Max}\left\{|u_{i}(t\!+\!\Delta t)\!-\!u_{i}(t)|\right\} \!<\! 1\!\times\! 10^{-7}, \\
			&{\rm Max}\left\{|v_{i}(t\!+\!\Delta t)\!-\!v_{i}(t)|\right\} \!<\! 1\!\times\! 10^{-7},\quad (1\!\leq\!i\leq\!N),\\
			& \Delta t\!=\!0.001. 
		\end{split}
	\end{eqnarray}
\end{enumerate}

Note that the convergent configurations are dependent on the initial configuration, which varies with random numbers except for $\vec{r}_i$ of the regular lattice sites. For this reason, to compute the mean values of the physical quantities, we repeat the steps (i)--(iv) by using different random numbers for the initial configurations. The mean value is calculated as follows: 
\begin{eqnarray}
	\label{mean-value}
	Q=\frac{1}{n_s}\sum_{i=1}^{n_s} Q_i,\quad n_s=200
\end{eqnarray}
where $Q_i$ is the lattice average of physical quantity $Q$ obtained from $i$-th convergent configuration produced in Step (iv). The number $n_s$ is the total number of samples.
These steps indicate that the TP system $(u,v)$ is influenced by the thermal fluctuations of $\vec{\tau}$ and $\vec{r}$ of membranes via the interactions with $\vec{\tau}$ and $\vec{r}$.

Fluctuations in the variables $(u,v)$ do not necessarily indicate that the system is in a non-equilibrium state. The non-equilibrium property is incorporated into the RD equation in Eq. (\ref{FN-eq-Eucl}) such that the equilibrium solutions of  $u_0$ and $v_0$ with $f\!=\!g\!=\!0$ change to $(u,v)$ under the reactions of $f$ and $g$. These non-trivial changes from $u_0$ to $u$ and $v_0$ to $v$ are accompanied by the energy transfers between $u$ and $v$, resulting in the solution $(u,v)$ of Eq. (\ref{FN-eq-Eucl}). Due to these implemented energy transfers, as described in the Introduction such that $\delta H\!=\!\delta (H_u\!+\!A H_v)\!=\!0$ even under $\delta H_{u}^D\!\not=\!0$ and $\delta H_{v}^D\!\not=\!0$, the steady state solution $(u,v)$ can be understood as a non-equilibrium steady state.

In contrast, the membrane system $(\vec{r}, \vec{\tau})$ is updated with the Metropolis MC to obtain the equilibrium state corresponding to the Hamiltonian in Eq. (\ref{discrete-total-Hamiltonian}). In this framework for $(\vec{r}, \vec{\tau})$,  it is not possible to obtain direct evidence of non-equilibrium states or processes such as relaxation, despite the observation of energy localisation in stretched membranes subjected to external mechanical forces.  Nevertheless, it will be demonstrated in the following Section that an approximate estimate of the relaxation of the polymer direction $\vec{\tau}$ can be made.

Note also that the TPs that emerge in Step (iii) are almost the same as those in the convergent configurations in Step (iv). From this convergence in Step (iii), we understand how  the TP system $(u,v)$ interacts with the membrane system  $(\vec{r}, \vec{\tau})$ in the context of hybrid simulations; $u,v$ are updated in the discrete time evolution of the RD equation, while $(\vec{r}, \vec{\tau})$ are updated in MC. The interference between $(u,v)$ and $(\vec{r}, \vec{\tau})$ is described in two steps. First, the $(u,v)$ are influenced by the $\vec{\tau}$ direction via the interaction implemented in the coefficients $D_{ij}^{u,v}(\vec{\tau})$ in the discrete RD equation.  Second, the direction $\vec{\tau}$ is determined via the coefficients $\Gamma_{ij}^{G,b}(\vec{\tau})$ in $H_{1}\!+\!\kappa H_2$ and also by the interaction energy $\lambda H_\tau$.

We comment on the lattice size dependence of TPs. Our simulation-checks indicate that the total number of TPs depends on the lattice size. If the numbers  $n_x$ and $n_y$ are increased by a factor of two will result in a total number of TPs that is twice as large. This phenomenon can be attributed to the fact that the spatial period of TPs,  or the distance between two TPs, depends mainly on the parameters $D_u, D_v$. An increase in the ratio $D_u/D_v$ results in an increase in the spatial period.  It is also necessary to modify the parameters $\alpha$ and $\gamma$ in order to accommodate a significant deviation of $D_u, D_v$ from the current values. For these reasons, the lattice size $N\!=\!2900$ is considered to be sufficiently large in this study \cite{note-6}.

\section{Results \label{Sec:result}}

\subsection{Turing patterns on membranes in 2-dim plane ${\bf R}^2$}
\begin{figure}[h!]
	\centering
	\includegraphics[width=8.5cm]{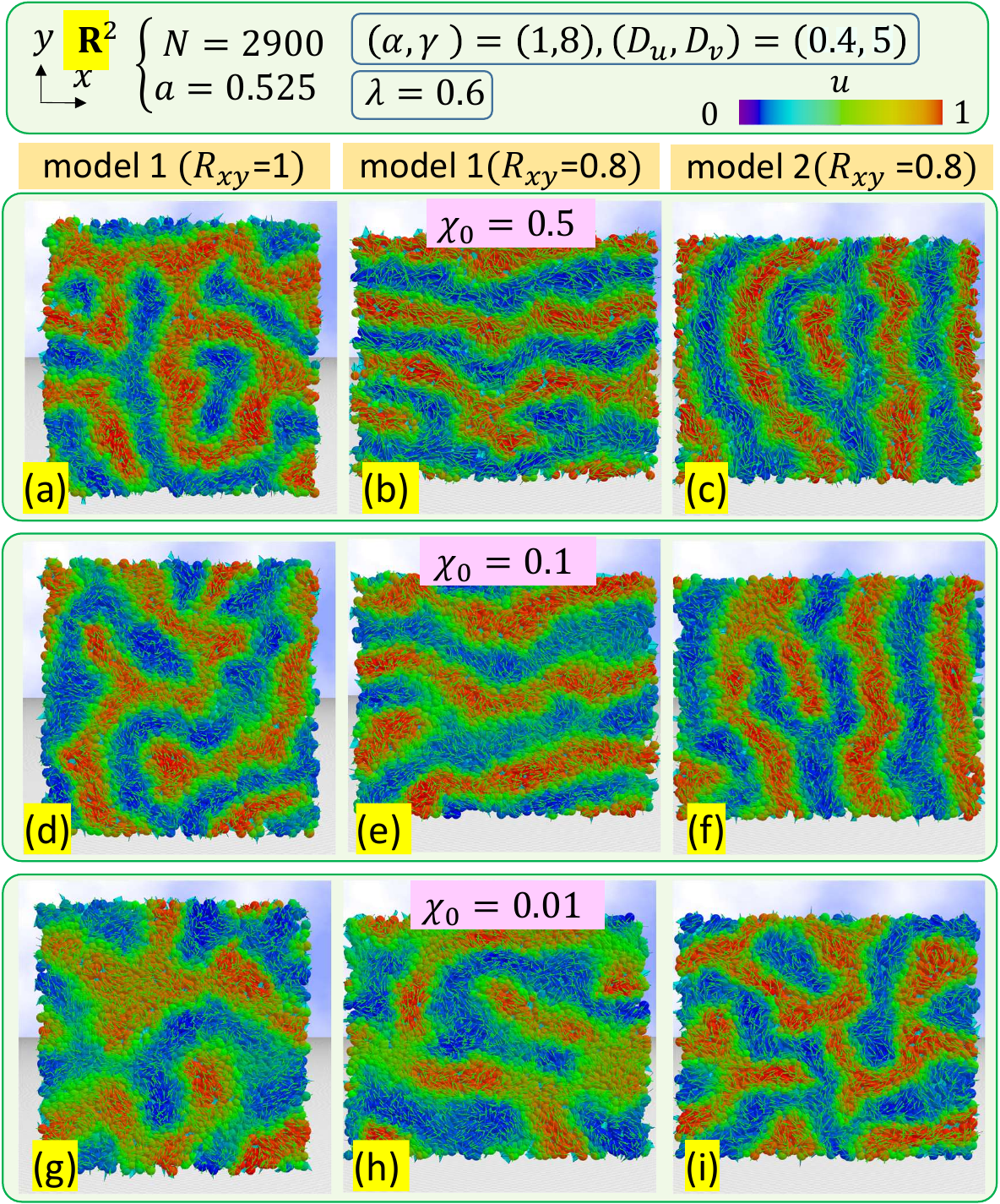}
	\caption{TPs in models 1,2 in ${\bf R}^2$ in response to stretching characterized by $R_{xy}$. Results of model 1 with $R_{xy}\!=\!1$, model 1 with $R_{xy}\!=\!0.8$,  and model 2 with $R_{xy}\!=\!0.8$, for (a),(b),(c) $\chi_0\!=\!0.5$, (d),(e),(f) $\chi_0\!=\!0.1$, and (g),(h),(i) $\chi_0\!=\!0.01$. The small cones represent $\vec{\tau}$, which is parallel to the TP direction in both models 1 and 2.  $\vec{\tau}$ is parallel (perpendicular) to the stretching direction in model 1 (model 2) for  $\chi_0\!=\!0.5$ and  $\chi_0\!=\!0.1$, while no strong correlation is observed for $\chi_0\!=\!0.01$. 
		\label{fig-7}}
\end{figure}
Now, we study TPs on stretched membranes. The parameter $\lambda$ is fixed to $\lambda\!=\!0.6$, where the surface entropy is maximized as shown below. We expect that the obtained surface configurations at $\lambda\!=\!0.6$ correspond to those in equilibrium in real membranes for any $R_{xy}$ close to $R_{xy}\!=\!1$.

First, we show that anisotropic TPs appear in models 1 and 2 in ${\bf R}^2$  in response to the surface stretching and to check whether TPs are the same as those in Ref. \cite{Diguet-etal-PRE2024}. The two-dimensional self-avoidance illustrated in Fig. \ref{fig-3}(a) is imposed as in the model in Ref. \cite{Diguet-etal-PRE2024}. The bending energy $H_2$ is not assumed because $H_2\!=\!0$ on the surface in ${\bf R}^2$. The models studied in this subsection are those obtained from models 1 and 2 in ${\bf R}^3$ by reducing the external space dimension to ${\bf R}^2$, where $\vec{\tau}$ has values in a half-circle; $\vec{\tau}\in S^1/2$ in ${\bf R}^2$. 

Three different values of $\chi_0$ in Eqs. (\ref{Finsler-unit-lengths-1}) and (\ref{Finsler-unit-lengths-2}) are examined such that
\begin{eqnarray}
	\label{chi-value-examin}
 {\rm (i)} \chi_0=0.5, \quad  {\rm (ii)}  \chi_0=0.1, \quad  {\rm (iii)} \chi_0=0.01. 	
\end{eqnarray}
Snapshots obtained at (i) $\chi_0\!=\!0.5$, (ii) $\chi_0\!=\!0.1$, and (iii) $\chi_0\!=\!0.01$ are shown in Figs. \ref{fig-7}(a),(b),(c),  Figs. \ref{fig-7}(d),(e),(f), and  Figs. \ref{fig-7}(g),(h),(i), respectively.

The presented snapshots illustrate that the TP direction exhibits characteristics of near-isotropy for $R_{xy}\!=\!1$ independent of $\chi_0$ (Figs. \ref{fig-7}(a),(d),(g)). In contrast, the TP direction  for $R_{xy}\!=\!0.8$  is parallel (Figs. \ref{fig-7}(b),(e)) and perpendicular (Figs. \ref{fig-7}(c),(f)) to the stretching direction at $\chi_0\!=\!0.5$ and $\chi_0\!=\!0.1$. In the case of $\chi_0\!=\!0.01$ (Figs. \ref{fig-7}(h),(i)), there is no clear correlation between the TP direction and the stretching direction.  Therefore, we find that the results of model 1 obtained at  $\chi_0\!=\!0.5$ and $\chi_0\!=\!0.1$ in Figs. \ref{fig-7}(a)(b) and \ref{fig-7}(d),(e) are consistent with those obtained in Ref. \cite{Diguet-etal-PRE2024}. The results of model 2 in Figs. \ref{fig-7}(c),(f),(i) could not be obtained with the vertex-move model of  Ref. \cite{Diguet-etal-PRE2024}, because $H_1$ in the model of Ref. \cite{Diguet-etal-PRE2024} is the standard Gaussian energy and the results are uniquely determined only by the stretching direction. This consistency between the results of model 1  and the model in Ref. \cite{Diguet-etal-PRE2024} implies that models 1 and 2 are well defined as extensions of the model presented in Ref. \cite{Diguet-etal-PRE2024}  at least for the range of $\chi_0\!=\!0.5$ and $\chi_0\!=\!0.1$. The TPs are independent of $\chi_0$ at the range of  $\chi_0\!=\!0.5$ and $\chi_0\!=\!0.1$, though $\chi_{ij}$ in Eqs. (\ref{Finsler-unit-lengths-1}) and (\ref{Finsler-unit-lengths-2}) becomes $\chi_{ij}\!\to\! \chi_0$ for sufficiently large $\chi_0$. 

From these examinations of the value of $\chi_0$, we assume the same value as in Ref. \cite{Diguet-etal-PRE2024} such that
\begin{eqnarray}
	\label{chi-value}
	\chi_0=0.5
\end{eqnarray}
in the simulations below for the models in ${\bf R}^2$ and ${\bf R}^3$. When the other value of  $\chi_0$ is used, it will be specified.

\subsection{Turing patterns on membranes in 3-dim space ${\bf R}^3$}
\begin{figure}[t]
	\centering
	\includegraphics[width=8.5cm]{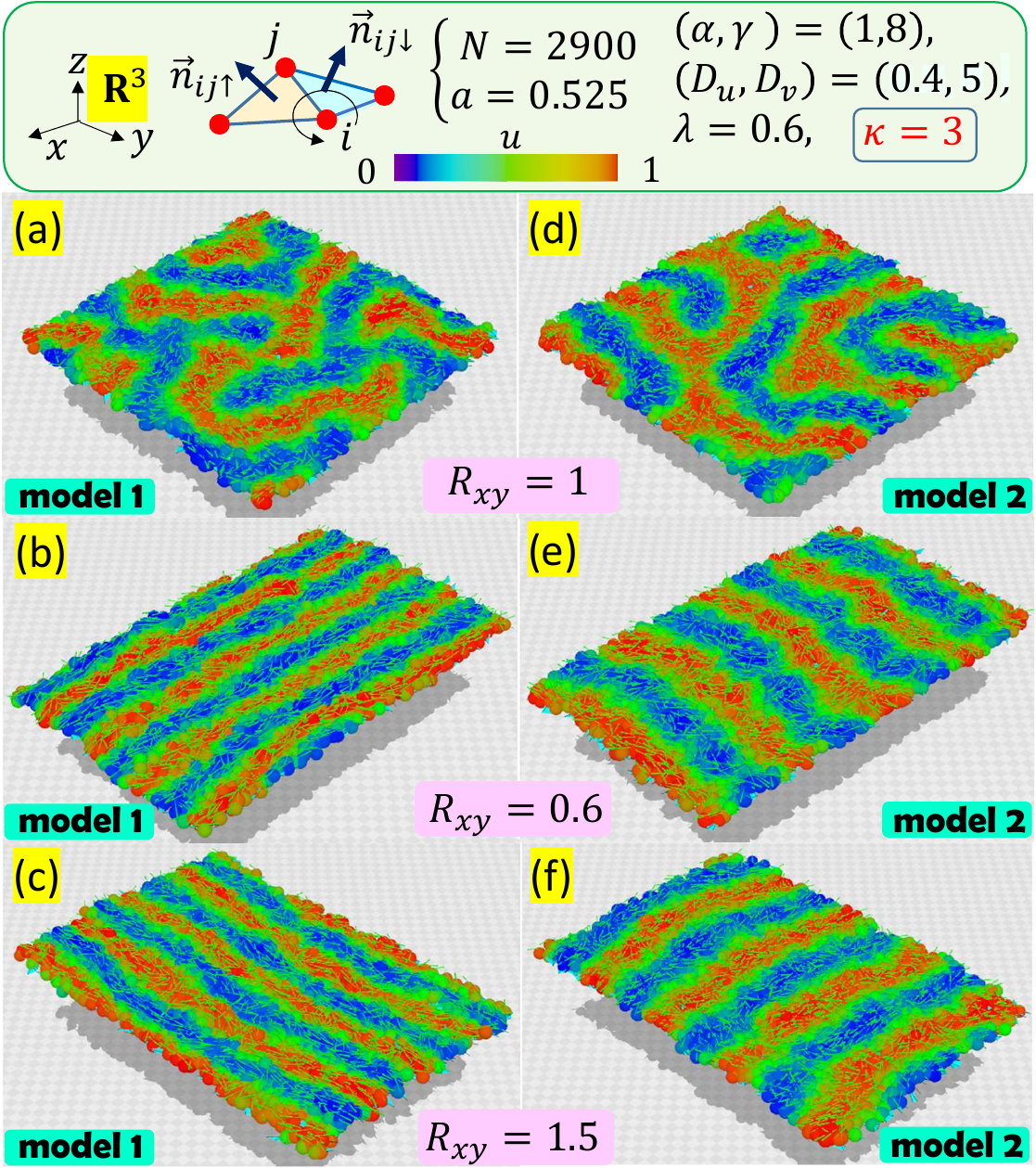}
	\caption{ The emergent TPs on membranes in ${\bf R}^3$ for $R_{xy}\!=\!1$, $R_{xy}\!=\!0.6$, $R_{xy}\!=\!1.5$ of (a), (b), (c) model 1, and (d), (e), (f) model 2, where $\kappa\!=\!3$. 
		\label{fig-8}}
\end{figure}
Next,  by including the bending energy $\kappa H_b$ in the Hamiltonian and letting $\vec{\tau}_i\in\! S^1/2$: half circle on the tangential plane at vertex $i$ (Fig. \ref{fig-4}(e)), we extend the space ${\bf R}^2$ to ${\bf R}^3$ and confirm that the properties of the models in ${\bf R}^2$ are preserved in ${\bf R}^3$. Given that the membranes span the boundary frame (Fig. \ref{fig-1}(a)), it is reasonable to expect that the non-self-intersecting property, or self-avoidance and surface smoothness, will be observed for sufficiently large values of $\kappa$. Snapshots obtained by models 1 and 2 at $\kappa\!=\!3$ are shown for $R_{xy}\!=\!1, 0.6, 1.5$ in Figs. \ref{fig-8}(a)--(f). The assumed parameters are shown on the figure, and these are used in all subsequent simulations except for $\kappa$ and $\lambda$, which are varied. We find that the TP directions that emerge in response to the surface stretching, including the case $R_{xy}\!=\!1$, are identical to those obtained on the plane ${\bf R}^2$ in Fig. \ref{fig-7}.  To  ascertain the effects of surface smoothness, we show the results obtained at $\kappa\!=\!1$ in Figs. \ref{fig-9}(a)--(f). The TP directions of model 2 in Figs. \ref{fig-9}(e),(f) at $\kappa\!=\!1$  are opposite to those of model 2 in Figs. \ref{fig-8}(e),(f) at $\kappa\!=\!3$.

\begin{figure}[th]
	\centering
	\includegraphics[width=8.5cm]{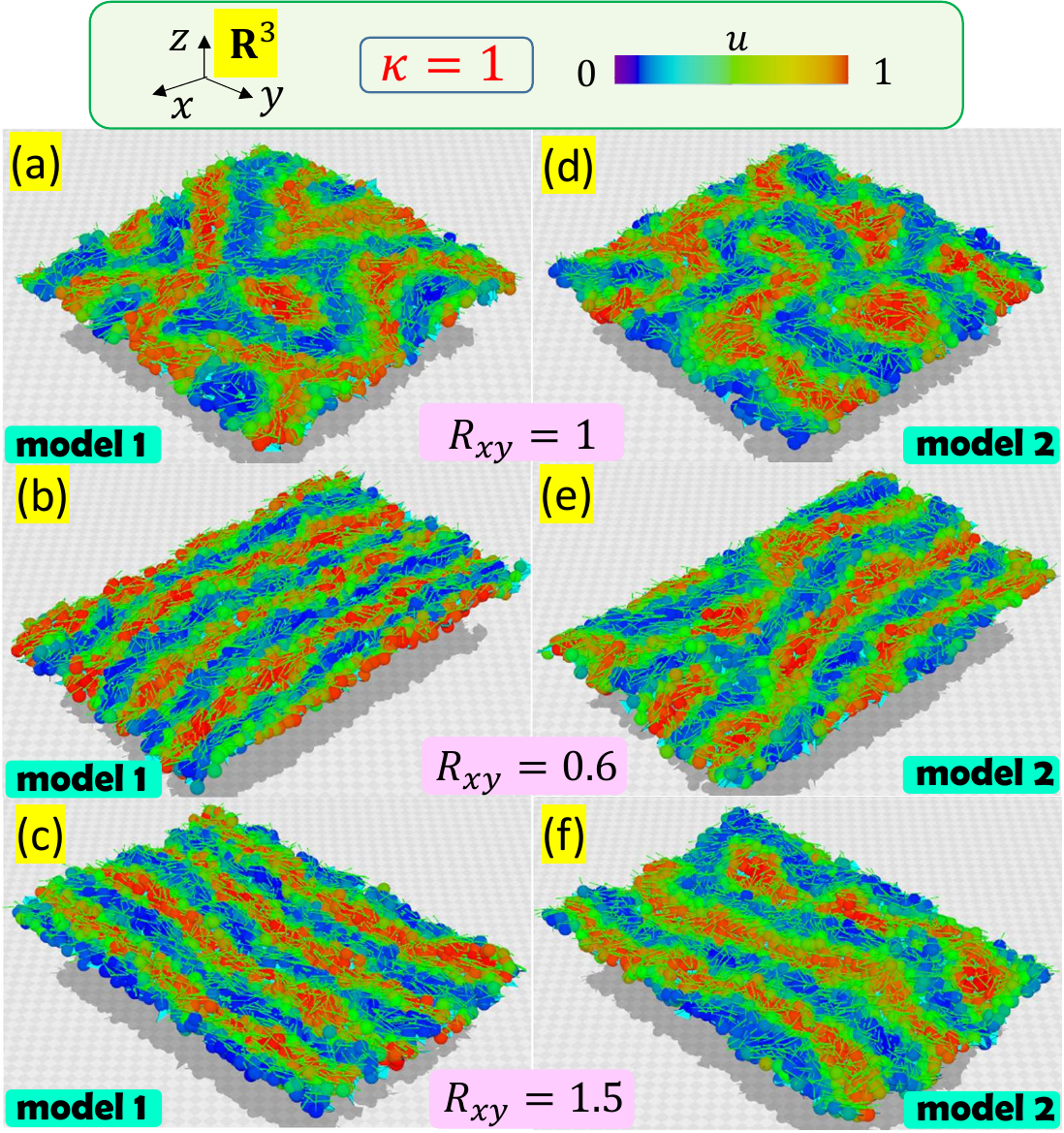}
	\caption{ Snapshots of stretched surfaces of model 1 (a),(b),(c) and model 2 (d),(e),(f) obtained at $\kappa\!=\!1$. TP directions of both models are along the stretched direction. Additionally, the surface roughness is relatively high and the TPs are less clear at $\kappa\!=\!1$ in comparison to those observed in the snapshots presented in Fig. \ref{fig-8}. 
		\label{fig-9}}
\end{figure}

\subsection{Maximum entropy state and relaxation of polymer structure under the membrane stretching}
It should first be noted that the argument presented in this subsection is contingent upon the inclusion of the IDOF $\vec{\tau}$ in the models. The relaxation of polymer structure is a time-dependent phenomenon and is not directly simulated in our MC technique used to update $(\vec{r}, \vec{\tau})$. However, due to the inclusion of $\vec{\tau}$ as a variable in the models for membrane, it is possible to obtain information on the relaxation from the entropy calculation.  As mentioned above,   the  strength of the interaction between $\vec{\tau}$ is dependent on  the coefficient  $\lambda$ of $\lambda H_\tau$, $\lambda$ plays a role in fixing the relaxation of the polymers.  In actual biological membranes, the internal structure, including the polymer direction, gradually reaches equilibrium configurations when subjected to stretching strains. These time-dependent configurations can be captured as equilibrium states in our modelling. To evaluate effects of variation of the polymer direction $\vec{\tau}$ on the entropy, the coefficient $\lambda$  associated with $H_\tau$ is varied in the simulations. Therefore, it can be postulated that the simulated steady-state configurations on stretched membranes at an arbitrary $\lambda$ correspond to a non-relaxed configuration involved in the process of relaxation.

\begin{figure}[h!]
	\centering
	\includegraphics[width=8.5cm]{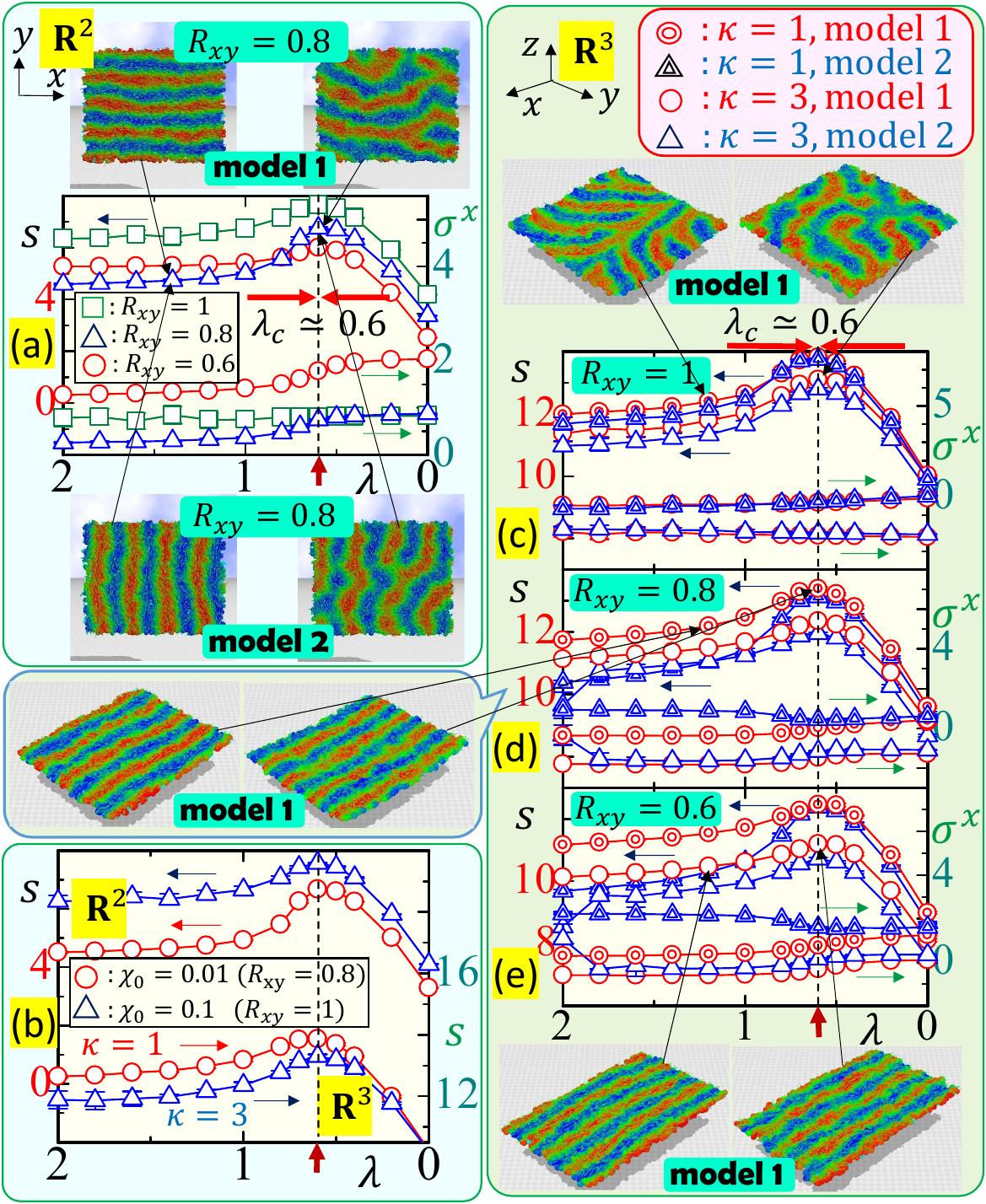}
	\caption{(a) Entropy $s(=\!\delta S/\delta  A_p)$ and surface tension $\sigma^x$ vs. $\lambda$  with snapshots for non-stretched ($R_{xy}\!=\!1$) and stretched ($R_{xy}\!\not=\!1$) membranes of model 1 in ${\bf R}^2$, including snapshots of model 2.  (b) The plots of $s$ vs. $\lambda$ for (\textcolor{red}{$\bigcirc$}$:\chi_0\!=\!0.01$, $R_{xy}\!=\!0.8$) and (\textcolor{blue}{$\triangle$}$:\chi_0\!=\!0.1$, $R_{xy}\!=\!1$) for model 1 in ${\bf R}^2$ and ${\bf R}^3$. $s$ and  $\sigma^x$ vs. $\lambda$ for (c) $R_{xy}\!=\!1$, (d) $R_{xy}\!=\!0.8$, and (e) $R_{xy}\!=\!0.6$  of models 1 and 2 in ${\bf R}^3$ with $\kappa\!=\!1$ and $\kappa\!=\!3$. The maximum entropy is observed universally at $\lambda_c\!\simeq\! 0.6$, where the relaxed or equilibrium state can be simulated for each $R_{xy}$. $\chi_0$ is fixed to $\chi_0\!=\!0.5$ except for (b).  \label{fig-10}}
\end{figure}
Figure \ref{fig-10}(a) shows the entropy $s(=\!\delta S/\delta  A_p)$ in Eq. (\ref{entropy-2-dim}) and the surface tension  $\sigma^x$ vs. $\lambda$ obtained on three different stretched membranes of $R_{xy}\!=\!1$, $R_{xy}\!=\!0.8$ and $R_{xy}\!=\!0.6$ for model 1 in ${\bf R}^2$.  Here we are interested in whether or not $s$ is increased by the stretching, so for simplicity the simulation unit, described by $k_BT\!=\!1$ and $a\!=\!0.525$, is used. 
Results $s$ of model 2 are similar to those in Fig. \ref{fig-10}(a) and are therefore not included in the plot. Snapshots include those of model 2. 

We find the maximum entropy state at $\lambda_c\!\simeq\! 0.6$, which is independent of $R_{xy}$. The dashed lines are drawn at  $\lambda\!=\!0.6$ as indicated by (\textcolor{red}{$\uparrow$}). The result $\lambda_c\!\simeq\! 0.6$ is also independent of $\chi_0$  at least for $\chi_0\!=\!0.01$ with $R_{xy}\!=\!0.8$ and $\chi_0\!=\!0.1$ with $R_{xy}\!=\!1$ for model 1 in ${\bf R}^2$ and ${\bf R}^3$, where $\kappa\!=\!1$ for $\chi_0\!=\!0.01$ and  $\kappa\!=\!3$ for $\chi_0\!=\!0.1$ (Fig. \ref{fig-10}(b)). Other physical quantities generally depend on $\chi_0$ in the models of ${\bf R}^2$ and ${\bf R}^3$.  For the  investigation in this paper, we have fixed $\chi_0\!=\!0.5$, as defined in Eq. (\ref{chi-value}), and will not be conducting further checks regarding the dependence of physical quantities on $\chi_0$. 
In the supplementary material (1), this problem is further discussed.  

The behaviour of $\lambda_c$ is worthy of further discussion.  To illustrate, if a configuration on the stretched membrane is identified with the state at $\lambda \!=\!1$, for example, then the configuration will be relaxed to the state at  $\lambda \!=\!\lambda_c$, as indicated by the arrows (\textcolor{red}{$\rightarrow\leftarrow$}) pointing toward to the dashed  line in Fig. \ref{fig-10}(a). In our models, configurations that are expected to emerge in time-dependent relaxation process can be simulated as equilibrium states by fixing $\lambda$ as  $\lambda\!\not=\!\lambda_c$. In this sense, the configuration of $(\vec{r},\vec{\tau})$ obtained for $\lambda\!\not=\!\lambda_c$ in our hybrid MC simulations corresponds to a non-equilibrium state that evolves towards the maximum entropy state in real membranes.  

Further results for the models in ${\bf R}^3$ are presented in Figs. \ref{fig-10}(c),(d),(e). The bending rigidity is fixed to two different values, $\kappa\!=\!3$ and $\kappa\!=\!1$, as assumed in Fig. \ref{fig-10}(b). We find that the maximum entropy position is $\lambda_c\!\simeq\!0.6$ for all $R_{xy}\!=\!1$, $R_{xy}\!=\!0.8$ and $R_{xy}\!=\!0.6$.  Including the results in Fig. \ref{fig-10}(b), we find that the $\lambda_c$  remains constant  irrespective of the specific model  and $\kappa$. Furthermore, it can be seen that $\lambda_c$ is identical to that observed in the model of  ${\bf R}^2$. The jumps of $\sigma^x$ at $\lambda\!=\!2$ in Figs. \ref{fig-10}(d),(e)  are a consequence of  the non-uniqueness of the TP direction. It should be noted that the Monte Carlo technique produces samples with TP configurations that differ from the expected minimum energy state or may be trapped in local minima. This is due to the value of  $\lambda(=\!2)$ being too large  for the configurations to be independent of those initially trapped. 
Detailed information on the entropy is presented in the supplementary material (2). 

The observed fact that the peak value $\lambda_c$ is not influenced by the variation of several parameters can be explained if the following condition is satisfied: the entropy variation is primarily influenced by the variation of $\lambda H_\tau$ with respect to the $\lambda$ variation. This condition is satisfied in $s$ in Eq. (\ref{entropy-2-dim}) for the models in ${\bf R}^2$, and it is also satisfied in $s$ in Eq. (\ref{entropy-3-dim}) for the models in ${\bf R}^3$ because the variation of $\kappa H_2$ is not expected by the $\lambda$ variation. However, it is crucial to emphasise that $\lambda_c\!\simeq\! 0.6$ is only valid  within a limited range of conditions that have been checked. For this reason, $\lambda_c\!\simeq\! 0.6$ plotted in Fig. \ref{fig-10} does not substantiate the existence of a universal value $\lambda_c$.

To summarise the results in this subsection with supplementary material (2): (i) the maximum entropy state is obtained at $\lambda\!\simeq\!0.6$, which is independent of the embedding dimension ${\bf R}^D, (D\!=\!2,3)$, model (1 or 2), in a limited range of  $R_{xy}$, $\kappa$ ($\kappa\!=\!1$ or $\kappa\!=\!3$ for the models in ${\bf R}^3$) and $\chi_0$, (ii) physically meaningful combinations in the choice of model and parameter ranges can be deduced from the surface tension calculation, (iii) non-equilibrium configurations can be simulated indirectly in the models with IDOF $\vec{\tau}$ by varying $\lambda$.

\section{Discussions on TP direction and membrane deformation \label{Sec:discussion}}
In this section, we discuss how the TP direction is determined in models 1 and 2. For this purpose, we introduce "easy axis  for deformation", which is a deformation direction with lower energy cost for  stretched membranes.

\subsection{Easy axes for tensile and bending deformations in the standard model for membranes}
 First, we should note that mechanical anisotropies such as tension anisotropy and bending anisotropy can be introduced on the stretched surfaces, which are described by the standard model for membranes defined by the Hamiltonian  \cite{KANTOR-NELSON-PRA1987,Gompper-Kroll-PRA1992,HELFRICH-1973,Polyakov-NPB1986,Peliti-Leibler-PRL1985,Bowick-PRep2001,NELSON-SMMS2004,Wheater-JPA1994,KOIB-PRE-2005}
\begin{eqnarray}
	\begin{split}
		\label{std-model}
		&H^{\rm std}(\vec{r})=H_1^{\rm std}+\kappa H_2^{\rm std},\\
		&H_1^{\rm std}\!=\!\sum_{ij}\ell_{ij}^2,\quad H_2^{\rm std}\!=\!\sum_{ij}(1\!-\!\vec{n}_{ij\downarrow}\cdot \vec{n}_{ij\uparrow}),
	\end{split}
\end{eqnarray}
where $H_1^{\rm std}$ and $H_2^{\rm std}$ are the bond potential and the bending energy corresponding to $H_1$ and $H_2$ in Eq. (\ref{discrete-Hamiltonian}). In the standard model for membranes, no variable for IDOF is included, and the membrane deformations are controlled only by  $\ell_{ij}^2$  and $1\!-\!\vec{n}_{ij\downarrow}\cdot \vec{n}_{ij\uparrow}$, which are the extensive parts of energies $H_1^{\rm std}$ and $\kappa H_2^{\rm std}$. The intensive parts in $H_1^{\rm std}$ and $\kappa H_2^{\rm std}$ are 1 and $\kappa$, indicating that no mechanical anisotropy is assumed for the stretching and bending. This assumption is valid for non-deformed standard membranes, however, such a mechanical isotropy is not always expected in the stretched membranes. 

\begin{figure}[h!]
	\centering
	\includegraphics[width=8.5cm]{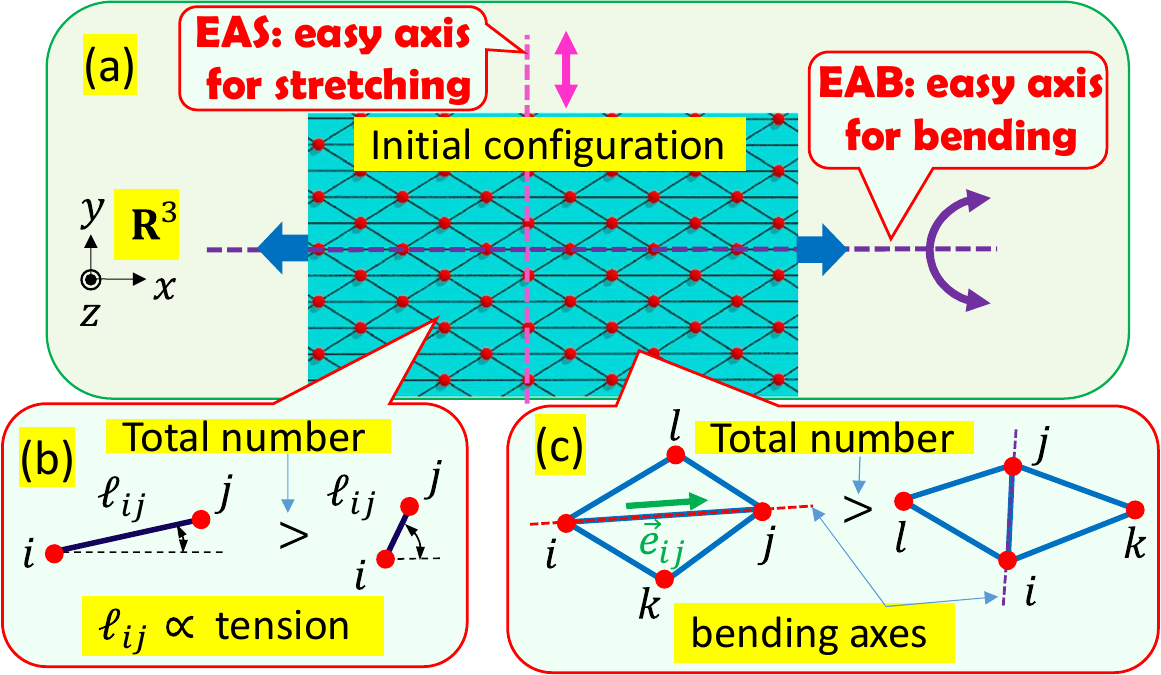}
	\caption{(a) A stretched initial surface configuration with the easy axes for stretching (EAS) and bending (EAB) in the standard model for membranes, (b) illustrations of the extended and shrunken bonds; the total number of the extended bonds (=almost parallel to the $x$-axis) is larger than that of the shrunken bonds (=almost vertical to the $x$-axis). (c) Illustrations of the bending axes parallel  to the $x$ and $y$ axes; the total number of the bending axes parallel to the $x$-axis is larger than that parallel to the $y$-axis. 
		\label{fig-11}}
\end{figure}
For a stretched surface shown in Fig. \ref{fig-11}(a), the $y$-axis is regarded as the "easy axis for stretching", due to the fact that the total number of stretched bonds along the $x$-axis is larger than that of the relatively short bonds along the $y$-axis (Fig. \ref{fig-11}(b)). In these stretched configurations, the tensile force for stretching along the $x$-axis is expected to be larger than that for stretching along the $y$-axis. Consequently, such stretched surfaces will have more bending axes along the $x$-axis rather than those along the $y$-axis (Fig. \ref{fig-11}(c)). This deformability for bending naturally leads to the notion of an "easy axis for bending" along the $x$-axis. 

Thus, we find that the concept of "easy axes for mechanical deformation" is well defined within the context of the standard triangulated surface model for membranes. We should note that the easy axes are introduced in an intuitive manner due to the fact that the tension coefficient $\gamma(\!=1)$ and the bending rigidity $\kappa$, which represent the intensive part of the energies, are constant and isotropic even at the microscopic level.

In contrast to the standard model, the easy axes for deformation are naturally introduced in FG models 1, 2 for membranes defined by $H_1$ and $H_2$ in Eq. (\ref{discrete-Hamiltonian}). In these models, the intensive part of energies, defined as $\Gamma_{ij}^{G,b}$, play a role in the strength of molecular forces at the microscopic level. Moreover, $\Gamma_{ij}^{G,b}$ vary depending on the configurations of $\vec{\tau}$, which can be controlled by external forces or strains. In other words, the FG model for membranes is suitable for implementing the mechanical anisotropies corresponding to stretching and bending deformations.

\subsection{Finsler geometry modelling of anisotropic membranes}
\subsubsection{Easy axes for stretching and bending}
Next, we discuss the easy axes in FG models 1 and 2 in ${\bf R}^2$, as well as on smooth surfaces in ${\bf R}^3$, where $\vec{\tau}$ aligns parallel (perpendicular) to the stretched direction in model 1 (model 2). Here, we employ the Hamiltonian $H\!=\!H_1\!+\!\kappa H_2$ to describe the mechanical property of membranes corresponding to the stretching and bending, and discard the other terms in Eq. (\ref{discrete-Hamiltonian}). In this subsection, based on the numerically obtained alignment directions of $\vec{\tau}$, we intuitively discuss that the easy axes for stretching and bending in FG models are the same as those in the standard model. We then proceed to introduce the definitions of the easy axes and related terminologies in the following subsection. 

\begin{figure}[h!]
	\centering
	\includegraphics[width=8.5cm]{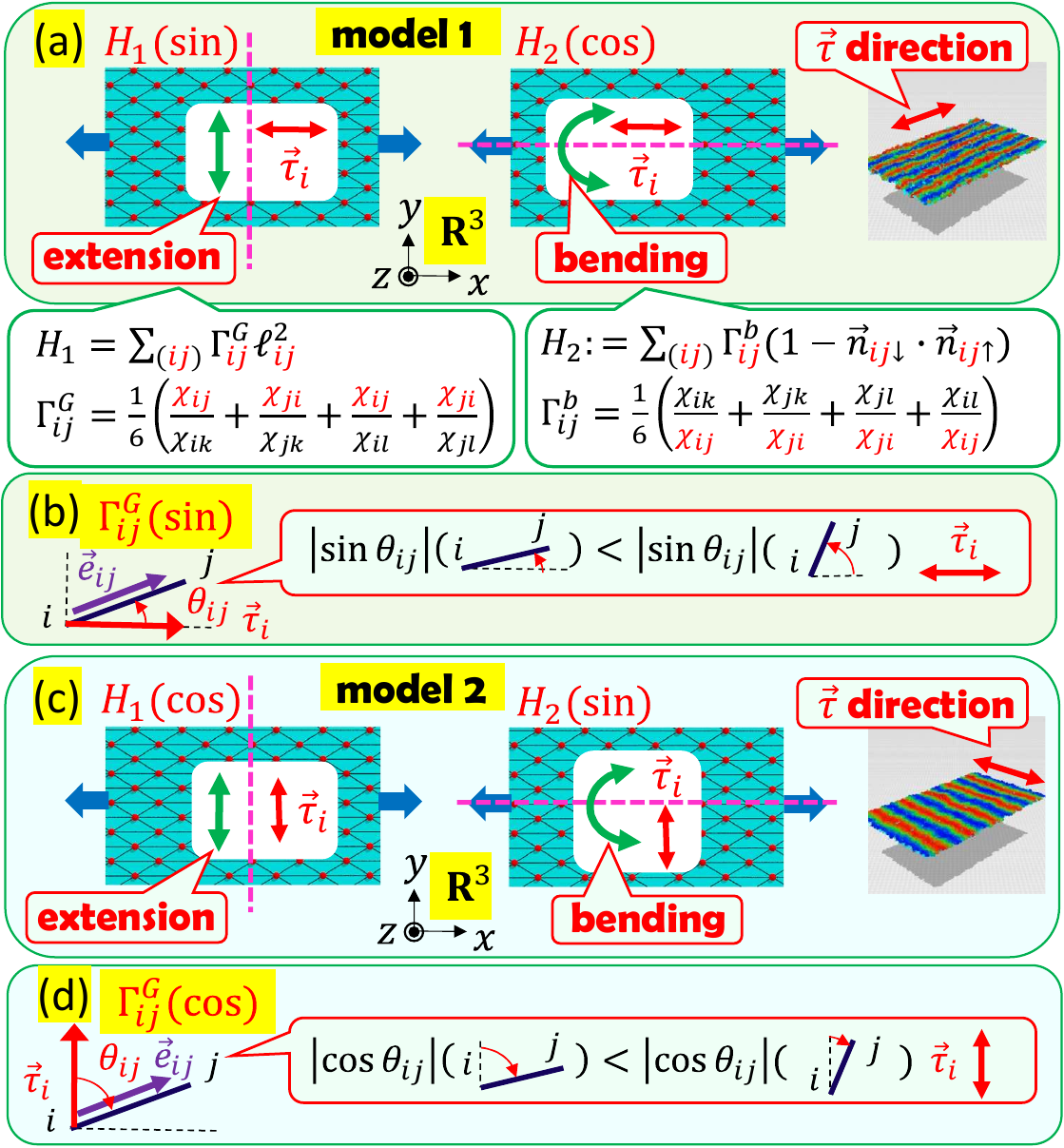}
	\caption{(a), (b) Easy axes for stretching and bending (dashed lines) expected in model 1 and model 2, where symbols sin and cos in $H_{1,2}$ denote $\chi_{ij}^{G,b}\!=\!\sqrt{1\!-\!|\vec{\tau}_i\cdot\vec{e}_{ij}|^2} \!+\! \chi_0 (\simeq\!\sqrt{1\!-\!|\vec{\tau}_i\cdot\vec{e}_{ij}|^2}\!=\!|\sin \theta_{ij}|)$ and $\chi_{ij}^{G,b}\!=\!|\vec{\tau}_i\cdot\vec{e}_{ij}| \!+\! \chi_0 (\!\simeq\!|\vec{\tau}_i\cdot\vec{e}_{ij}|\!=\!|\cos \theta_{ij}|)$, respectively. $\Gamma_{ij}^G$ ($\Gamma_{ij}^b$) includes $\chi_{ij}$ and $\chi_{ji}$ in the numerators (denominators). (a),(b) Due to $|\sin \theta_{ij}|$ from $\chi_{ij}$ in $H_1({\rm sin})$ for model 1, $\Gamma_{ij}^G$ on bond $ij$ parallel to the stretched direction is smaller (larger) than that perpendicular to the stretched direction when $\vec{\tau}$ is parallel (perpendicular) to the stretched direction. (c),(d) In $H_1({\rm cos})$ for model 2, the same results are obtained. \label{fig-12}}
\end{figure}
Figures \ref{fig-12}(a)--(d) show surfaces stretched in the $x$ direction with information on $\Gamma_{ij}^{G,b}$ from the FG model Hamiltonian.   The coefficient $\Gamma_{ij}^G$ along bond $ij$  depends on $\chi_{ij}$ or $\chi_{ji}$ (Eqs. (\ref{Finsler-unit-lengths-1}) and (\ref{Finsler-unit-lengths-2})), which are proportional to $|\sin \theta_{ij}| (\simeq\! |\sin\theta_{ji}|)$ for model 1 and $|\cos \theta_{ij}| (\simeq\! |\cos\theta_{ji}|)$ for model 2, where $\chi_0(>\!0)$ is the common additive number and is neglected for simplicity.  From the expression of $\Gamma_{ij}^G$ for model 1 with $\vec{\tau}$ parallel to the $x$-axis, we naturally expect that  $\Gamma_{ij}^G$ is larger on bonds along the $y$-axis than that along the $x$-axis (Fig. \ref{fig-12}(b)). This large value of $\Gamma_{ij}^G$ implies a small $\ell_{ij}^2$ along the $y$-axis. In other words, $\Gamma_{ij}^G$ on the stretched bond $ij$ along the $x$-axis  becomes small while the bond length becomes large. This results in a large energy $\Gamma_{ij}^G\ell_{ij}^2$  along the stretched $x$ direction and is consistent with the configurations forced by the stretching. Due to this large tensile energy along the stretched direction, further stretching along the $y$-axis will be easier than along the $x$-axis. Thus, the easy axis for stretching is identified as the $y$-axis in model 1.

In contrast, the expression of the coefficient $\Gamma_{ij}^b$ in $H_2^b$  differs from that of $\Gamma_{ij}^G$; $\chi_{ij}$ and $\chi_{ji}$ are included in the denominators of $\Gamma_{ij}^b$ (Fig. \ref{fig-12}(a)), moreover,  $\Gamma_{ij}^b$ is the cosine type. For these reasons, $\Gamma_{ij}^b$  on bond $ij$ along the $y$-axis is also larger than on bond $ij$ along the $x$-axis,  implying that the easy axis for bending is the $x$-axis. Thus, the easy axes for stretching and bending are expected to be the same as those in the standard model plotted in Fig. \ref{fig-9}(a). It should be noted  that these discussions are limited to moderately deformed surfaces, where the extensive part of the energies  $\ell_{ij}^2$  and $1\!-\!\vec{n}_{ij\downarrow}\cdot \vec{n}_{ij\uparrow}$ are only slightly different from those of the unstretched  surface.

In the case of model 2,  $\vec{\tau}$ is numerically shown to be parallel to the $y$-axis, as illustrated in Fig. \ref{fig-12}(c). In model 2, $\Gamma_{ij}^G$ and $\Gamma_{ij}^b$ are the cosine and sine types, respectively,  and thus these are larger along the $y$-axis than those along the $x$-axis (Fig. \ref{fig-12}(d)). Consequently, the same discussions, applied to model 1 in Figs. \ref{fig-12}(a),(b), can be applied to model 2.  

To summarize the discussions for stretched surfaces along the $x$ direction ($\Leftrightarrow R_{xy}\!<\!1$);
\begin{eqnarray}
	\label{discussion-summary}
	\begin{split}
	&	\Gamma_{ij}^G \; (\vec{e}_{ij} \parallel \vec{e}^{\,x} ) <  \Gamma_{ij}^G \; (\vec{e}_{ij} \parallel  \vec{e}^{\,y} ) \quad({\rm models \; 1,\; 2})\\
	&	\Gamma_{ij}^b \; (\vec{e}_{ij} \parallel \vec{e}^{\,x} ) <  \Gamma_{ij}^b \; (\vec{e}_{ij} \parallel  \vec{e}^{\,y} )  \quad({\rm models \; 1,\; 2})
	\end{split}
\end{eqnarray}
under the conditions:  
\begin{enumerate}
\item[(a)] $\vec{\tau}$ parallel to the $x$-axis ($y$-axis) in model 1 (model 2), 
\item[(b)] $R_{xy}(<\!1)$ is close to 1, or in other words, $\vec{e}_{ik}$, $\vec{e}_{il}$, $\vec{e}_{jk}$ and $\vec{e}_{jl}$ are not parallel to $\vec{e}_{ij}$ (Fig. \ref{fig-11}(c))), 
\end{enumerate}
where   $\vec{e}^{\,x}$ and  $\vec{e}^{\,y}$ are the unit vectors along the $x$- and $y$-axes.  Note that the out-of-plane surface fluctuation is also assumed to be small in the discussions for the models in ${\bf R}^3$. 

The aforementioned discussions are based on the assumption that the direction of $\vec{\tau}$ is parallel (perpendicular) to the stretched direction in model 1 (model 2) as shown in the snapshots in  Figs. \ref{fig-12}(a) and (c). In the opposite case that the $\vec{\tau}$ direction is perpendicular (parallel) to the stretched direction in model 1 (model 2), the expected directional energy distribution and easy axes are incompatible with those of the configurations forced by the extension.  This energetically inconsistent case is expected in the models in ${\bf R}^2$ for $\chi_0\!=\!0.01$ (see the supplementary material (1)).

\subsubsection{Easy axes and directional system}
First, we rewrite $H_1$ and $H_2$ as  (Appendix \ref{App-D})
\begin{eqnarray}
	\begin{split}
		\label{dirdipendent-H1H2}
		&H_1(\vec{r},\vec{\tau})=\Gamma_x^G H_1^x + \Gamma_y^G H_1^y,\\
		&H_2(\vec{r},\vec{\tau})=\Gamma_x^b H_2^x + \Gamma_y^b H_2^y. 
	\end{split}
\end{eqnarray}
For surfaces in ${\bf R}^2$, the first expression is sufficient, and the second one is unnecessary. For surfaces in ${\bf R}^3$,  the remaining terms $\Gamma_z^{G} H_1^z$ and $\Gamma_z^{b} H_2^z$ are necessary. However, the surface is nearly flat and parallel to the $xy$ plane, and thus, the additional terms are omitted for simplicity in the following discussion. The direction-dependent coefficients $\Gamma_\mu^{G,b}, (\mu\!=\!x,y)$ in Eq. (\ref{dirdipendent-H1H2}) are defined by
\begin{eqnarray}
	\label{effective-surface-tension}
	\begin{split}
	&\Gamma^G_x(\vec{r},\vec{\tau})=\frac{1}{\sum_{ij}|\vec{e}_{ij}^{\;x}|}\sum_{ij}\Gamma^G_{ij}|\vec{e}_{ij}^{\;x}|, \\
	&\Gamma^G_y(\vec{r},\vec{\tau})=\frac{1}{\sum_{ij}|\vec{e}_{ij}^{\;y}|}\sum_{ij}\Gamma^G_{ij}|\vec{e}_{ij}^{\;y}|, 
	\end{split}
\end{eqnarray}
where  $H_{1,2}^\mu,  (\mu\!=\!x,y)$ are the direction-dependent Gaussian bond potential and bending energy,  respectively. 

In the expressions $\Gamma^G_\mu(\vec{r},\vec{\tau})$ in Eq. (\ref{effective-surface-tension}), $|\vec{e}_{ij}^{\;\mu}|$ represents the $\mu$ component of the unit vector $\vec{e}_{ij}$, and it is defined in Eq. (\ref{direction-dependent-Gamma}). The factor $|\vec{e}_{ij}^{\;\mu}|$ is independent of $\vec{\tau}$ and reflects the effects of lattice deformation by the stretching. Consequently,  $\Gamma^G_\mu(\vec{r},\vec{\tau})$ are influenced by the direction of bond $ij$ in two ways: firstly, by $\Gamma^G_{ij}$ as discussed in the preceding subsection, and secondly, by $|\vec{e}_{ij}^{\;\mu}|$. These two effects   influencing  $\Gamma^G_\mu(\vec{r},\vec{\tau})$ depend on  $\chi_0$.  Further details can be found in the supplementary material (1).

To define the easy axis for stretching (bending), it is sufficient to evaluate which of $H_1^x$ and $H_1^y$ ($H_2^x$ and $H_2^y$ ) is smaller or larger. If $H_1^x$ is larger than $H_1^y$ for a stretched surface,  we understand that the easy axis for stretching is the $y$-axis due to the lower energy cost for stretching in the $y$ direction. Thus, we define
\begin{eqnarray}
	\label{easy-axes-definition}
	\begin{split}
		&y{\rm-axis}={\rm easy \; axis\;for\; stretching}\; ({\rm EAS}) \Leftrightarrow H_1^x > H_1^y,\\
		&x{\rm-axis}={\rm easy \; axis\;for\; bending}\;  ({\rm EAB}) \Leftrightarrow H_2^x > H_2^y.
	\end{split}
\end{eqnarray}
If the symbol "$>$" is replaced by "$<$" on the right hand side, the easy axis is the vertical one to that described on the left hand side.

We should note that each term on the right hand side of Eq. (\ref{dirdipendent-H1H2}) is positive definite and well defined for Hamiltonian describing a physical system,   and hence we call the system "the directional system" of membranes described by $H^x$ and $H^y$ such that
\begin{eqnarray}
	\label{directional-system}
	\begin{split}
		&	H(\vec{r},\vec{\tau})=H_1+\kappa H_2= H^x+H^y,\\
		&H^x(\vec{r},\vec{\tau})=\Gamma_x^G H_1^x\!+\!\kappa \Gamma_x^b H_2^x,\quad H^y(\vec{r},\vec{\tau})=\Gamma_y^G H_1^y\!+\!\kappa \Gamma_y^b H_2^y.
	\end{split}
\end{eqnarray}
$H_{1,2}^{\mu}, (\mu=x,y)$ play a role in the Gaussian bond potential and the bending energy along the $\mu$ direction. 
These systems emerge when the membranes are stretched and disappear at $R_{xy}\!\to\! 1$ such that 
\begin{eqnarray}
	\label{directional-system-reduce}
	\begin{split}
		&\Gamma_x^G=\Gamma_y^G\to\Gamma_0^G,\quad \Gamma_x^b=\Gamma_y^b\to\Gamma_0^b, \\
		&H_1^x=H_1^y\to H_1^0,\quad H_2^x=H_2^y\to H_2^0, \quad (R_{xy}\to 1),
	\end{split}
\end{eqnarray}
where $\Gamma_0^G$, $\Gamma_0^b$, $H_1^0$ and $H_2^0$ denote the coefficients and energies at $R_{xy}\to 1$ and are the same as those of the isotropic membranes up to the irrelevant multiplicative factor $1/2$.

	\subsubsection{Stretched membranes and energy localisation}
It should be noted that the extensive part of the energies is determined by the intensive parts $\Gamma_\mu^{G,b}, (\mu\!=\!x,y)$, and vice versa in the directional systems, such that  
\begin{eqnarray}
	\label{equilibrium-H12}
	\begin{split}
		&\Gamma_x^{G}(< \Gamma_0^{G}) < \Gamma_y^{G} \Leftrightarrow H_{1}^x(>H_{1}^0) > H_{1}^y,\\
		&\Gamma_x^{b}(< \Gamma_0^{B}) < \Gamma_y^{b} \Leftrightarrow H_{2}^x(>H_{2}^0) > H_{2}^y.
	\end{split}
\end{eqnarray}
Note that these are one of the possible conditions satisfied in the configurations close to the isotropic stable state at $R_{xy}\!=\!1$ corresponding to $\delta H_1\!=\!\delta(\Gamma_x ^{G}H_{1}^x\!+\!\Gamma_y ^{G}H_{1}^y)\!\!=\!0$, $\delta H_2\!=\!\delta(\Gamma_x ^{b}H_{2}^x\!+\!\Gamma_y ^{b}H_{2}^y)\!=\!0$ indicated by the condition (i) in Eq. (\ref{two-combinations}) (Appendix \ref{App-D}). 
	Another possible condition for the energy localization corresponding to (ii) in Eq. (\ref{two-combinations}) is
\begin{eqnarray}
	\label{equilibrium-H12-another}
	\begin{split}
		&\Gamma_x^{G}(> \Gamma_0^{G}) > \Gamma_y^{G} \Leftrightarrow H_{1}^x(>H_{1}^0) > H_{1}^y,\\
		&\Gamma_x^{b}(> \Gamma_0^{B}) > \Gamma_y^{b} \Leftrightarrow H_{2}^x(>H_{2}^0) > H_{2}^y.
	\end{split}
\end{eqnarray}

The conditions in  Eqs. (\ref{equilibrium-H12}) and (\ref{equilibrium-H12-another}) imply an emergence of the directional system.  To define the directional energy transfer from the $y$ to $x$ directions in a directional system, both the intensive and extensive parts of energies are used such that
	\begin{eqnarray}
		\label{equilibrium-H12-extensive}
		\begin{split}
			& {\rm Tensile\; energy}: y \to x \Leftrightarrow \Gamma_x^G H_{1}^x(>\Gamma_0^G H_{1}^0) > \Gamma_y^G H_{1}^y, \\
			&  {\rm Bending\; energy}: y \to x  \Leftrightarrow \Gamma_x^b H_{2}^x(>\Gamma_0^b H_{2}^0) > \Gamma_y^b H_{2}^y.
		\end{split}
	\end{eqnarray}
We should note that the energy configurations  at $R_{xy}\!=\!1$ satisfy $\Gamma_x^G H_{1}^x\!=\!\Gamma_y^G H_{1}^y$ and $\Gamma_x^b H_{2}^x \!=\! \Gamma_y^b H_{2}^y$ for the tensile and bending energies.  When $R_{xy}$ deviates from  $R_{xy}\!=\!1$, the energy transfers in Eq. (\ref{equilibrium-H12-extensive}) are expected. This expectation is numerically confirmed below.

  \subsection{Numerical data}
 
 \subsubsection{Elastic energy localisation for stretching and bending of the models in ${\bf R}^2$}
 \begin{figure}[h!]
	\centering
	\includegraphics[width=8.5cm]{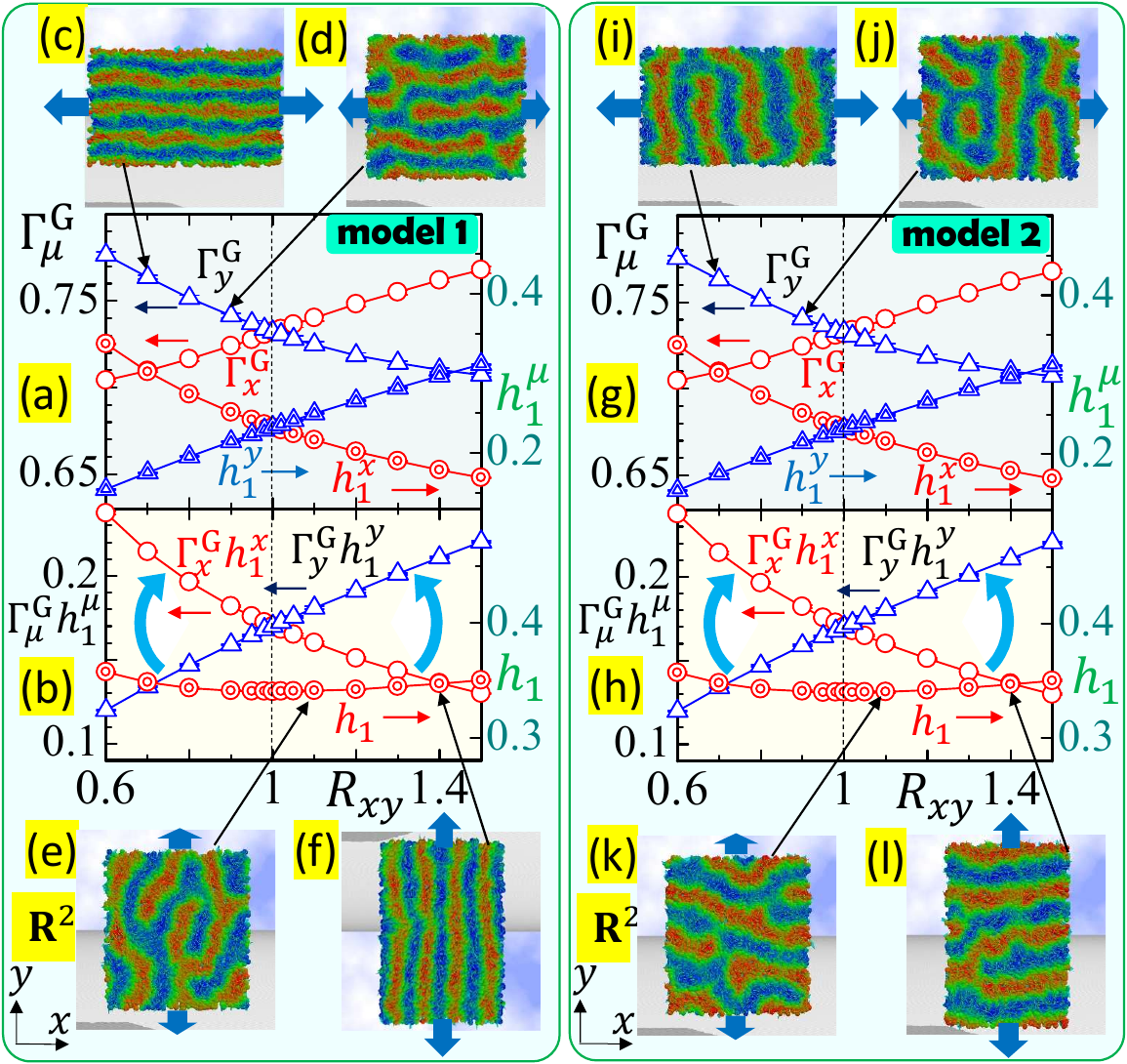}
	\caption{The direction-dependent surface tension $\Gamma^G_\mu, (\mu\!=\!x,y)$,  the direction-dependent energies $h_1^\mu, (\mu\!=\!x,y)$ and  $\Gamma^G_\mu h_1^\mu, (\mu\!=\!x,y)$  vs. $R_{xy}$ with snapshots of  (a)--(f)  model 1 and (g)--(l) model 2 for membranes in ${\bf R}^2$. $N\!=\!2900$ with $n_x\!=\!50$, $n_y\!=\!58$. The easy axis for stretching parallel (perpendicular) to the TP directions in model 1 (model 2) is consistent with the expectation in the preceding subsection. The energy transfer ($\Leftrightarrow \Gamma^G_x h_1^x\!\not=\!\Gamma^G_y h_1^y$) is observed in (b), (h) as indicated by the curved arrows  $\circlearrowright$ and $\circlearrowleft$. 
 \label{fig-13}}
\end{figure}
To numerically show that the definition of the easy axes for stretching in Eq. (\ref{easy-axes-definition}) is well defined, we plot the direction-dependent coefficient $\Gamma_{ij}^G$  for surface tension and the corresponding energy 
\begin{eqnarray}
	\label{G-energy-per-bond}
h_1^\mu=H_1^\mu/N_B, (\mu=x,y), \quad N_B=\sum_{ij}1=3N 
\end{eqnarray}
with snapshots of models 1 and 2 for membranes in ${\bf R}^2$, where the bending energy $H_2$ is not included (Figs. \ref{fig-11}(a)--(l)). 
The direction-dependent $\Gamma_x^G$ of model 1 (Fig. \ref{fig-13}(a)) is defined, as in Eq. (\ref{effective-surface-tension}), by the sum of small $\Gamma_{ij}^G$ for $R_{xy}<\!1$ as summarized in Eq. (\ref{discussion-summary}), and the resulting $\Gamma_x^G$ decreases (increases) with decreasing $R_{xy}(<\!1)$ (increasing $R_{xy}(>\!1)$).
In contrast, the behaviour of $\Gamma_y^G$ is opposite to that of $\Gamma_x^G$. In model 2, the behaviours of $\Gamma_\mu^G, (\mu\!=\!x,y)$  (Fig. \ref{fig-13}(g)) are almost the same as in model 1, because the coefficient $\Gamma_{ij}^G$ is cosine type; $H_1(\cos)$, and the $\vec{\tau}$ direction is perpendicular to the stretched direction  \cite{note-7}.

From the direction-dependent energies $h^x_1$ and $h^y_1$ of model 1 in Fig. \ref{fig-13}(a), we find that  $h^x_1\!>\!h^y_1$ for $R_{xy}\!<\!1$ and $h^x_1\!<\!h^y_1$ for $R_{xy}\!>\!1$. The results imply that the easy axis for stretching is the $y$-axis for $R_{xy}\!<\!1$ and the $x$-axis for $R_{xy}\!>\!1$ from Eq. (\ref{easy-axes-definition}). Thus, we confirm that the easy axis for stretching is vertical to the stretched direction along which TP aligns in model 1. In model 2, we also confirm from $h^x_1$ and $h^y_1$ in Fig. \ref{fig-13}(g) that  the easy axis for stretching is vertical to the stretched direction along which TP perpendicularly aligns. 

The energy transfer, described by $\Gamma_\mu^G h_1^\mu$ in  Eq. (\ref{equilibrium-H12-extensive}), is independent of the models as confirmed in Figs. \ref{fig-13}(b),(h). In the remainder of this paper, we will examine whether or not this independence is confirmed in ${\bf R}^3$ using the data $\Gamma_\mu^{G,b}$ and  $h_{1,2}^\mu$.

\subsubsection{Elastic energy localisation for stretching and bending of the models in ${\bf R}^3$}
 \begin{figure}[h!]
 	\centering
 	\includegraphics[width=8.5cm]{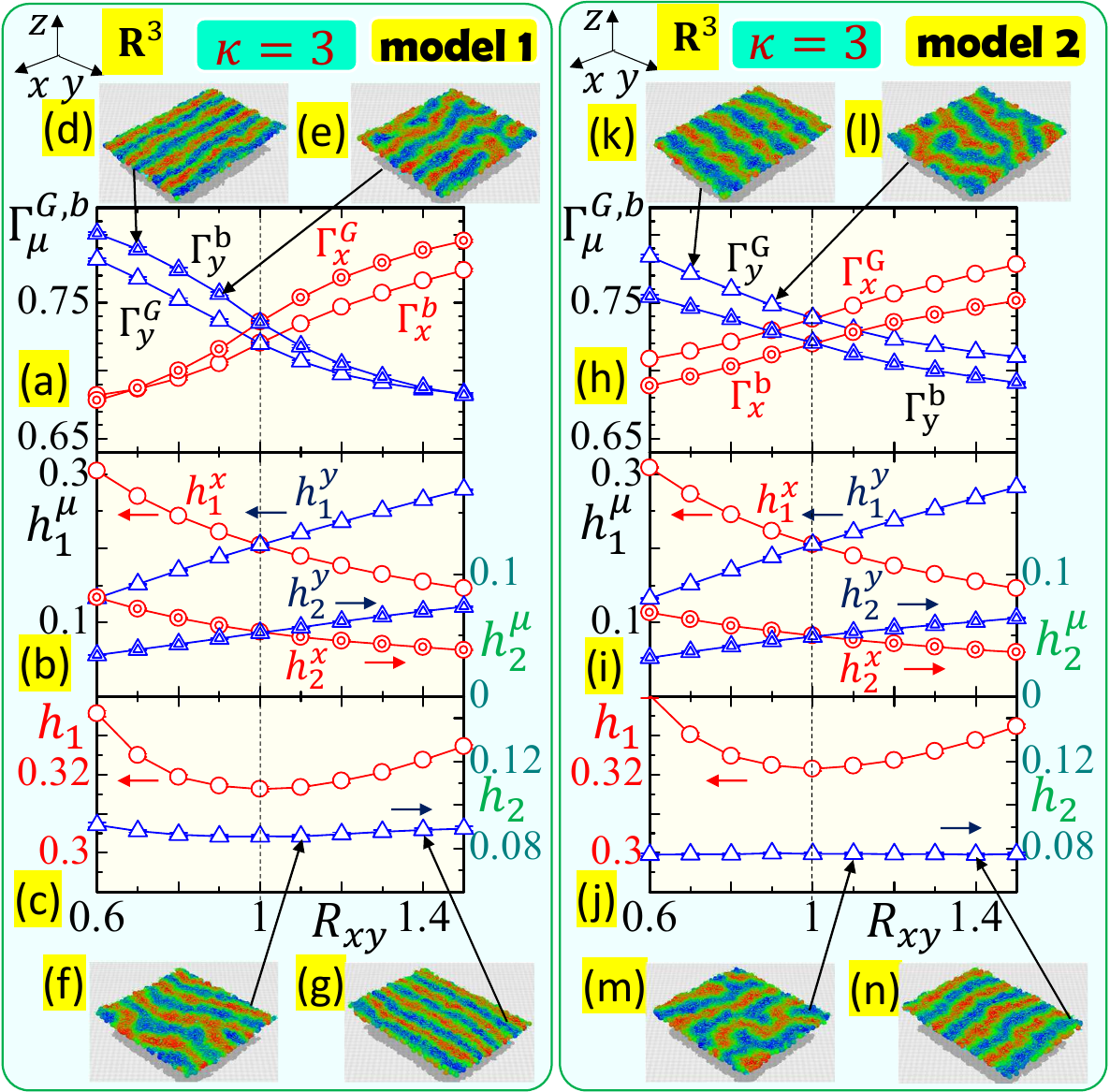}
 	\caption{The direction-dependent coefficients $\Gamma_\mu^{G,b}$, the direction-dependent energies $h^{G,b}_\mu$, and $h_1(=\!H_1/N_B)$ and $h_2(=\!H_2/N_B)$  with snapshots of (a)--(g) model 1 and (h)--(n)  model 2 for membranes in ${\bf R}^3$ with $\kappa\!=\!3$. $N\!=\!2900$ with $n_x\!=\!50$, $n_y\!=\!58$.  Note that the observation $h^x_{2}\!>\!h^y_{2}$ for $R_{xy}\!<\!1$ ($h^x_{2}\!<\!h^y_2$ for $R_{xy}\!>\!1)$) means a larger surface bending along the $x$-axis ($y$-axis) than that along the $y$-axis ($x$-axis) and is intuitively reasonable. \label{fig-14}}
 \end{figure}
Now, we plot the results of models 1 and 2 in ${\bf R}^3$ for $\kappa\!=\!3$ in Figs. \ref{fig-14}(a)--(n), where $\Gamma_\mu^b, (\mu\!=\!x,y)$ and $h_2^\mu, (\mu\!=\!x,y)$ for the bending energy are included. We find that the results of model 1 in Fig. \ref{fig-14}(a) are close to those of model 1 in Figs. \ref{fig-13}(a) in ${\bf R}^2$. We should note that $\Gamma_\mu^b$ vary with $R_{xy}$ almost the same as those of $\Gamma_\mu^G$. We also find from Fig. \ref{fig-14}(b) that $h^x_{1,2}\!>\!h^y_{1,2}$ for $R_{xy}\!<\!1$ and $h^x_{1,2}\!<\!h^y_{1,2}$ for $R_{xy}\!>\!1$ indicating that the easy axes for stretching and bending are consistent with the expectations discussed using Figs. \ref{fig-12}(a),(b), and find that the definitions in Eq. (\ref{easy-axes-definition}) are well defined.

The relations in Eq. (\ref{equilibrium-H12-extensive}) for $\Gamma_\mu^{G,b}h^\mu_{1,2}$ can also be checked using the data $\Gamma_\mu^{G,b}$ and $h^\mu_{1,2}$, and the energy transfers in the directional system are similar to those in the ${\bf R}^2$ models in Figs. \ref{fig-13}(a),(b). The results $h_1(=\!H_1/N_B)$ and $h_2(=\!H_2/N_B)$ in Fig.\ref{fig-14}(c) vary smoothly and remain almost unchanged at $R_{xy}\!\to\!1$. The results of model 2 shown in Figs. \ref{fig-14}(h)--(n) are almost the same as those of model 1 except for the TP direction. These observations indicate that the relationship between the TP direction and the directional energy configurations observed in the models in ${\bf R}^2$ remains unchanged when the models are extended to those in ${\bf R}^3$, in which the bending energy is included with a large bending rigidity $\kappa\!=\!3$. 

\begin{figure}[h!]
	\centering
	\includegraphics[width=8.5cm]{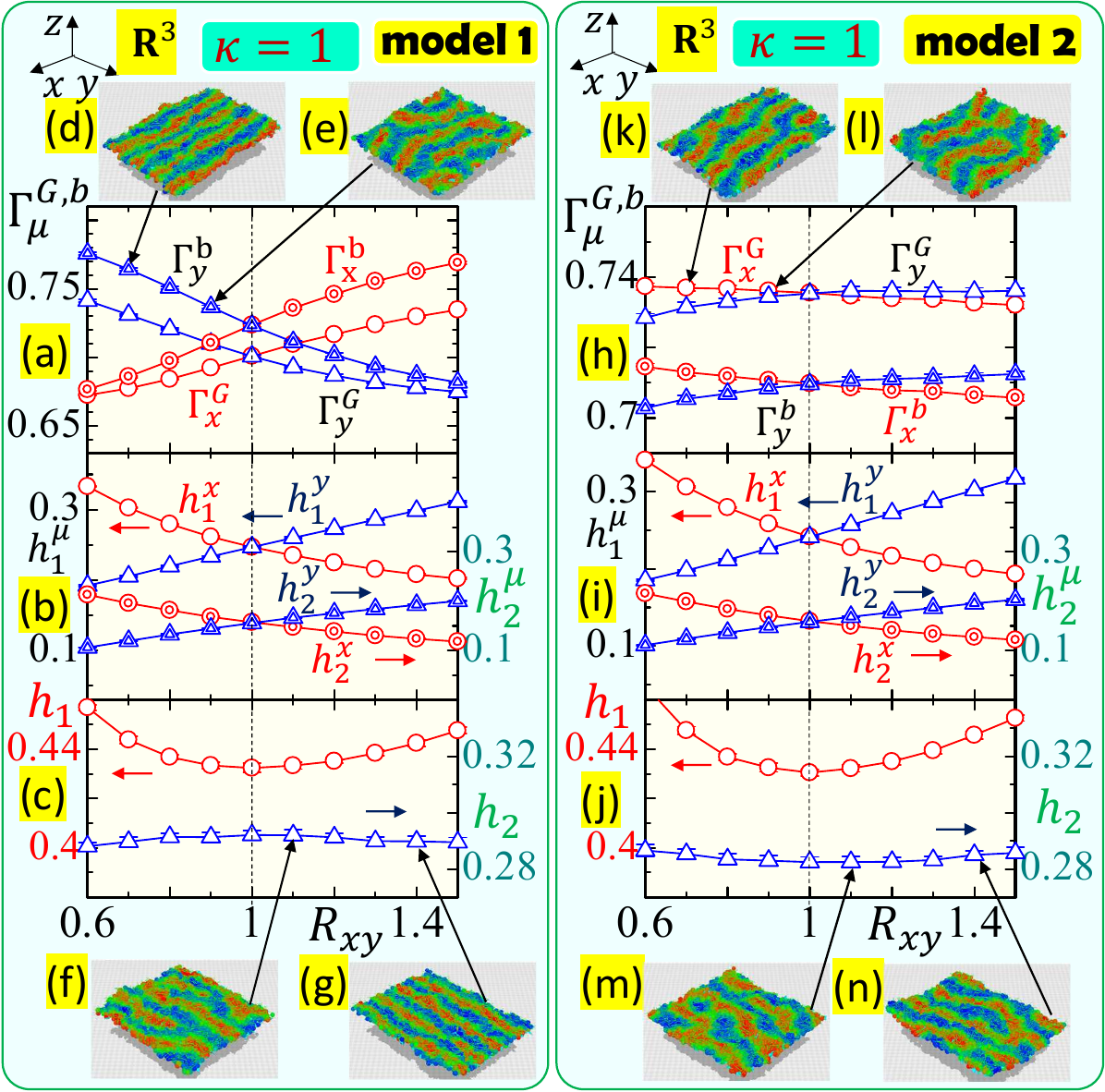}
	\caption{ $\Gamma_\mu^{G,b}$,  $h_{1,2}^\mu, (\mu\!=\!x,y)$,  and  $h_1$ and $h_2$ with snapshots of (a)--(g) model 1 and (h)--(n)  model 2 for membranes in ${\bf R}^3$ with $\kappa\!=\!1$. $\Gamma_\mu^{G,b}$ and  $h_{1,2}^\mu$  in (a), (b)  satisfy the relations in Eq. (\ref{equilibrium-H12}), while those in (h), (i) satisfy the relations in Eq. (\ref{equilibrium-H12-another}), 
		and hence, the data from both model 1 and model 2 meet the energy transfer conditions in Eq. (\ref{equilibrium-H12-extensive}). \label{fig-15}}
\end{figure}
The data of models 1 and 2 for $\kappa\!=\!1$ are shown in Figs. \ref{fig-15}(a)--(n).  It can be observed that  $\Gamma_x^{G,b}\!<\!\Gamma_y^{G,b}$ and $h^x_{1,2}\!>\!h^y_{1,2}$ for $R_{xy}\!<1\!$  in Figs. \ref{fig-15}(a),(b), and  $\Gamma_x^{G,b}\!>\!\Gamma_y^{G,b}$ and $h^x_{1,2}\!>\!h^y_{1,2}$ for $R_{xy}\!<1\!$ in Figs. \ref{fig-15}(h),(i). These observed conditions satisfy the relations in Eqs. (\ref{equilibrium-H12}) and (\ref{equilibrium-H12-another}), respectively. Thus, the easy axes for stretching and bending and the energy transfer directions of models 1 and 2 for  $\kappa\!=\!1$ are the same as those for $\kappa\!=\!3$.

\begin{table}[h!]
	\caption{\ The easy axes for stretching (EAS) and bending (EAB) and the TP directions for models 1 and 2 in ${\bf R}^3$ at $\kappa\!=\!3$ and  $\kappa\!=\!1$. The symbol $\leftrightarrow$ ($\updownarrow$) denotes the $x$-axis ($y$-axis).}
	\label{table-1}
	\begin{tabular*}{0.48\textwidth}{@{\extracolsep{\fill}}cccccc}
		\hline
		model & $\kappa$ & $R_{xy}$ & EAS (stretch) & EAB (bend) & TPs \\
		\hline
		1 &  3 &  $R_{xy}<1$ & {$\updownarrow$} & {$\leftrightarrow$} & {$\leftrightarrow$} \\
		2 &  3 & $R_{xy}<1$ & {$\updownarrow$} & {$\leftrightarrow$} & {$\updownarrow$} \\
		\hline
		1 &  1 & $R_{xy}<1$ & {$\updownarrow$} & {$\leftrightarrow$} & {$\leftrightarrow$} \\
		2 &  1 & $R_{xy}<1$ & {$\updownarrow$} & {$\leftrightarrow$} & {\textcolor{red}{$\leftrightarrow$}} \\
		\hline
	\end{tabular*}
\end{table}
In Table \ref{table-1}, easy axes (EAS, EAB) and TP directions of models 1 and 2 in ${\bf R}^3$ are listed. The symbols $(\updownarrow)$ and $(\leftrightarrow)$ denote that the direction is parallel to the $y$-axis and the $x$-axis, respectively. The results for $R_{xy}\!>1\!$ are opposite to those for $R_{xy}\!<1\!$ and are not listed.  We should note that EAS, EAB and TPs of the models for $\kappa\!=\!1$ are the same as those for $\kappa\!=\!3$, except for TPs of model 2 (\textcolor{red}{$\leftrightarrow$} and $\updownarrow$). The results of EAS and the TP direction of the models in ${\bf R}^2$ are the same as those for the models in ${\bf R}^3$ for  $\kappa\!=\!3$.

In supplementary material (3), the results of the diffusion energy localisation corresponding to $H_u$ and $H_v$ are presented, and the relationship between the TP direction and the diffusion directions of the chemical reactants $u$ and $v$ is discussed.

\section{Conclusions}
This paper presents a numerical study of Turing patterns (TPs) on fluctuating surfaces spanning a square frame with periodic boundary conditions in both two-dimensional plane (${\bf R}^2$) and three-dimensional space (${\bf R}^3$). This study is based on a triangulated surface model for membranes, which is defined by using the Finsler geometry (FG) modelling technique with an internal degree of freedom (IDOF) corresponding to the polymer directions. The FG modelling with the same IDOF is also used to define the anisotropic diffusion in the TP system. In this approach, the TPs on membranes are considered as a combined system comprising both the Turing system of activators/inhibitors and the membranes themselves. When the surface is subjected to stretching, the easy axes of stretching and bending are naturally introduced into the membrane sector. To find the response of TPs to the membrane deformation, two distinct models are  examined; model 1 and model 2. In these models, the implemented Finsler lengths for mechanical anisotropies are in opposition to one another as a response to the polymer direction, allowing the verification of direction-dependent interaction coefficients and energies, which can then be employed in the analysis of the relationship between the TP direction and the easy axes.

We find that the TP direction depends on the mechanical properties of the membranes under stretching. These properties are characterized by the easy axes for stretching and bending. The easy axis for stretching is perpendicular to that for bending in both models 1 and 2, and the easy axes in model 1 are identical to those in model 2. The easy axes are independent of the bending rigidity $\kappa$, which is related to the surface smoothness, in both models. With regard to  the relationship between the TP direction and the easy axes,  a numerical investigation of model 1 reveals  that the TP direction is perpendicular (parallel) to the easy axis for stretching (bending). In contrast, the TP direction in model 2 is found to be antiparallel to the easy axes for sufficiently large $\kappa$. To be more specific, the TP directions of model 1 are independent of $\kappa$ and uniquely determined by the easy axes for stretching and bending. In contrast, the TP directions of model 2 are dependent on $\kappa$. 

Furthermore, it has been demonstrated that non-equilibrium steady states can be simulated indirectly using the canonical Monte Carlo simulation.  This is due to the fact that the external force-induced anisotropy is correctly implemented in the FG modelling prescription in such a way that the anisotropy is effectively determined by the polymer direction.  It is evident that the stretching of membranes alters the mechanical energy distribution from uniform to directional, thereby resulting in a dynamical localisation of the energy distribution. This spatial energy rearrangement leads to a reduction in the thermodynamic entropy of the stretched membranes. Therefore, by varying the coefficient $\lambda$ of the correlation energy of the polymer directions $\vec{\tau}$, we identify the maximum entropy point $\lambda_c$  for a limited range of parameters and conditions. The existence of $\lambda_c$ indicates that the configurations obtained at $\lambda\!\not=\!\lambda_c$ under stretching correspond to non-equilibrium states in the process of relaxation of the polymer structure in real stretched membranes. 

Further numerical simulations are required to elucidate the relaxation problem. On a regular square lattice, the lattice expansion along the $x$-axis causes the variation  to be confined to the extensive part of the Gaussian bond potential. It is interesting to study the maximum entropy condition with respect to the variation of $\lambda$ on the regular square lattice and on the 3D materials in which the thickness can be varied with the expansion. 

\section*{Author Contributions}
F.K.: Project administration, Funding acquisition. H.K.: Methodology discussions, Simulations, Writing - Original Draft. E.B. C.C. R.D. S.M. M.N. and S.T. Methodology discussions, Writing - Review and Editing. T.U.: Supervision,  Resources. All authors discussed the techniques and results, approved the final version of the manuscript.

\section*{Data Availability Statement}
The data that support the findings of this study are available from the corresponding author upon reasonable request.

\section*{Conflicts of interest}
 There are no conflicts to declare. 

\section*{Acknowledgements}
This work is supported in part by Collaborative Research Project J24Ly01 of the Institute of Fluid Science (IFS), Tohoku University, and JSPS KAKENHI Grant Number JP23K03225, JP23K03208. Numerical simulations were performed on the supercomputer system AFI-NITY under the project CL01JUN24 of the Advanced Fluid Information Research Center, Institute of Fluid Science, Tohoku University.


\appendix

\section{Hamiltonians for Turing patterns} \label{App-A}
Starting with the continuous Hamiltonian 
\begin{eqnarray}
	\label{Continuous-Hamil-for-TP}
	\begin{split}
		&H=H_u+AH_v,\\
		&H_u=\int \sqrt{g}d^2x \left(\frac{1}{2}D_ug^{ab}\frac{\partial u}{\partial x^a}\frac{\partial u}{\partial x^b}-[\frac{u^2}{2}-\frac{u^4}{4}-Buv]\right),\\
		&H_v=\int \sqrt{g}d^2x \left(\frac{1}{2}D_vg^{ab}\frac{\partial v}{\partial x^a}\frac{\partial v}{\partial x^b}-\gamma[Buv-\alpha \frac{v^2}{2}-\beta v]\right),
	\end{split}
\end{eqnarray}
we show that the non-zero parameters $A$ and $B$ are uniquely determined for the Turing equation such that
\begin{eqnarray}
	\label{parameters-ABC}
		A=-\frac{1}{\gamma}(<0),\quad B=\frac{1}{2}.
\end{eqnarray}
 $H_{u,v}$ include the reaction terms corresponding to $f$ and $g$ in Eq. (\ref{FN-eq-Eucl}).   By the variational technique with respect to $u$ and $v$, we have
\begin{eqnarray}
	\label{variation-u-v}
	\begin{split}
	&0=\delta H=\frac{\delta H}{\delta u}\delta	u+\frac{\delta H}{\delta v}\delta v\\ 
	&=\left(\frac{\delta H_u}{\delta u}+A\frac{\delta H_v}{\delta u}\right)\delta u+\left(\frac{\delta H_u}{\delta v}+A\frac{\delta H_v}{\delta v}\right)\delta v \\
    &=\int \sqrt{g}d^2x \left(D_ug^{ab}\frac{\partial \delta u}{\partial x^a}\frac{\partial u}{\partial x^b}-[u-u^3-Bv]\delta u-A\gamma B v\delta u\right) \\
    &+A\int \sqrt{g}d^2x \left(A^{-1}Bu\delta v+D_vg^{ab}\frac{\partial \delta v}{\partial x^a}\frac{\partial v}{\partial x^b}-\gamma[Bu-\alpha v-\beta]\delta v\right)
	\end{split}
\end{eqnarray}
for arbitrary variations $\delta u$ and $\delta v$. Thus, we obtain
\begin{eqnarray}
	\label{variation-u-v-2}
	\begin{split}
		&-D_u \frac{1}{\sqrt{g}}\frac{\partial}{\partial x^a}\left(\sqrt{g}g^{ab}\frac{\partial  u}{\partial x^b}\right)-\left(u-u^3-[B-AB \gamma]v\right)=0, \\
		&-AD_v \frac{1}{\sqrt{g}}\frac{\partial}{\partial x^a}\left(\sqrt{g}g^{ab}\frac{\partial  v}{\partial x^b}\right)-A\gamma \left([B-\frac{B}{A\gamma}]u-\alpha v-\beta\right)=0. \\
	\end{split}
\end{eqnarray}
By letting  $B\!-\!AB \gamma\!=\!1$ and $B\!-\!\frac{B}{A\gamma}\!=\!1$, we find $A\!=\!-1/\gamma$ and $B\!=\!1/2$, and therefore
\begin{eqnarray}
	\label{FN-eq-curved-surface}
	\begin{split}
		&D_u \frac{1}{\sqrt{g}}\frac{\partial}{\partial x^a}\left(\sqrt{g}g^{ab}\frac{\partial  u}{\partial x^b}\right)+\left(u-u^3-v\right)=0, \\
		&D_v \frac{1}{\sqrt{g}}\frac{\partial}{\partial x^a}\left(\sqrt{g}g^{ab}\frac{\partial  v}{\partial x^b}\right)+\gamma \left(u-\alpha v-\beta\right)=0, 
	\end{split}
\end{eqnarray}
which are identical with the steady-state RD equation in Eq. (\ref{FN-eq-Eucl}) for $\beta\!=\!0$. 

We should note that the negative factor $A\!=\!-1/\gamma$  of $H_v$ in the linear combination of $H\!=\!H_u\!-\!\frac{1}{\gamma}H_v$ is a source of the interference between the two distinct states $u$ and $v$. As a consequence of the negative factor, the condition of $\delta H\!=\!0$ is compatible with $\delta H_u\!>\!0$ and  $\delta H_v\!>\!0$. It thus follows that the solution $(u,v)$ of Eq. (\ref{FN-eq-curved-surface}) is not a trivial extension of the constant functions $(u_0,v_0)$ of the diffusion equations that satisfy the conditions   $\delta H_u^D\!=\!0$ and $\delta H_v^D\!=\!0$. From an energetic  standpoint, the zero diffusion energies, represented by $H_u^D(u_0,v_0)\!=\!0$ and $H_v^D(u_0,v_0)\!=\!0$,  are transformed into non-zero values such that $H_u^D(u,v)\!>\!0$ and $H_v^D(u,v)\!>\!0$, due to the interactions corresponding to the second terms in Eq. (\ref{FN-eq-curved-surface}). Let 
\begin{eqnarray}
\begin{split}
&H_u^D(u,v)=H_u^D(u_0,v_0)+\delta H_u^D=\delta H_u^D,\\
&H_v^D(u,v)=H_v^D(u_0,v_0)+\delta H_v^D=\delta H_v^D,
\end{split}
\end{eqnarray}
 then the resulting situation is thus described as  $H_u^D(u,v)$ increasing by $\delta H_u^D$, while $-\!\frac{1}{\gamma}H_v^D$ decreases by $\frac{1}{\gamma}\delta H_v^D$. Therefore, we observe an "energy transfer" from $H_v^D(u,v)$ to $H_u^D(u,v)$. Consequently, the configuration change from the initial state $(u_0,v_0)$ to the current state $(u,v)$, which is accompanied by an energy transfer, is regarded as a transition from the isotropic equilibrium state to the anisotropic TP state.  This transition is independent of the membrane stretching. Furthermore, the directional diffusion energy transfer is also expected on the stretched membranes. This problem is discussed in supplementary material (3).

\section{Discrete Hamiltonians} \label{App-B}
\begin{figure}[h!]
	\centering
	\includegraphics[width=6.5cm]{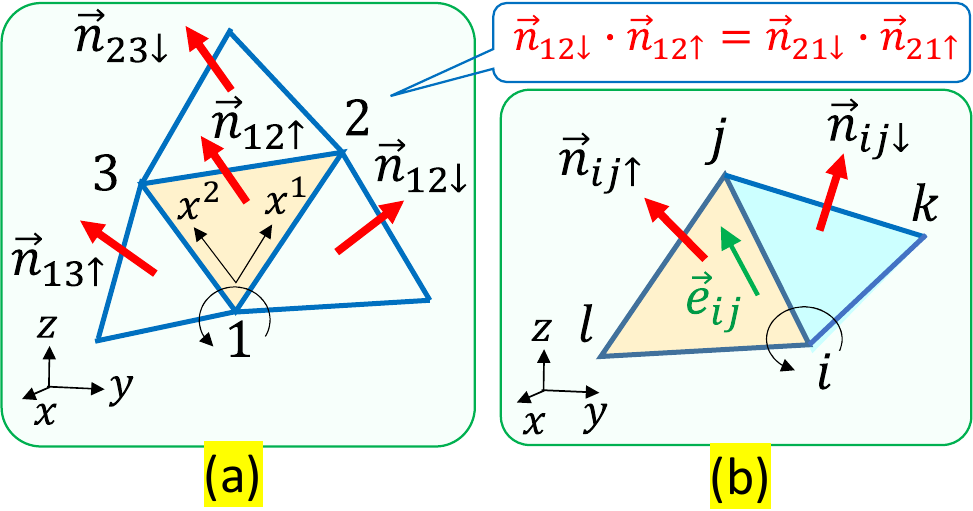}
	\caption{(a) Triangle 123 and the neighbouring triangles with the normal vectors indicated by arrows, (b) triangles sharing bong $ij$ and the normal vectors. \label{fig-B-1}}
\end{figure}
In this Appendix, we show how to obtain the discrete Hamiltonians from the continuous ones described by Finsler metrices, of which the discrete forms are described by 
\begin{eqnarray}
	\label{Finsler-metric-App}
	g^G= 
	\begin{pmatrix}
		1/(\chi_{ij}^G)^{2} & 0 \\ 
		0 & 1/(\chi_{ik}^G)^{2} 
	\end{pmatrix}, 
	\quad
	g^b= 
	\begin{pmatrix}
		1/(\chi_{ij}^b)^{2} & 0  \\
		0 & 1/(\chi_{ik}^b)^{2} 
	\end{pmatrix}
\end{eqnarray}
for $H_1$ and $H_2$, respectively, where $\chi_{ij}^{G,b}$ are given in Eqs. (\ref{Finsler-unit-lengths-1}), (\ref{Finsler-unit-lengths-2}). The metrices $g_{ab}^{u,v}$ for $H_u$ and $H_v$ are given in Appendix B in Ref.  \cite{Diguet-etal-PRE2024} and are skipped. Here, we only describe the discretization of $H_2$, for simplicity: 
\begin{eqnarray}
	\label{Continuous-bending-energy}
		H_2=\frac{1}{2}\int \sqrt{g}d^2x g^{ab}\frac{\partial \vec{n}}{\partial x^a}\cdot\frac{\partial \vec {n}}{\partial x^b}.
\end{eqnarray}
The continuous Hamiltonian $H_2$ is descretized on the triangle $123$ in Fig. \ref{fig-B-1}(a), where the local coordinate origin is at vertex 1, and $x^1$ and $x^2$ are the coordinate axes.   Replacing 
\begin{eqnarray}
	\label{discrete-form-1}
	\begin{split}
&\int \sqrt{g}d^2x\to \sum_{\Delta}\frac{1}{\chi_{12}^b\chi_{13}^b}, \\
&\frac{\partial \vec{n}}{\partial x^1}\to\vec{n}_{13\downarrow}-\vec{n}_{13\uparrow},\quad
\frac{\partial \vec{n}}{\partial x^2}\to\vec{n}_{12\uparrow}-\vec{n}_{12\downarrow},
	\end{split}
\end{eqnarray}
we have
\begin{eqnarray}
	\label{discrete-form-2}
	\begin{split}
 H_2\rightarrow &(1/2)\sum_{\triangle}\left(\frac{\chi_{12}^b}{\chi_{13}^b}(\vec{n}_{13\downarrow}-\vec{n}_{13\uparrow})^2+\frac{\chi_{13}^b}{\chi_{12}^b}(\vec{n}_{12\uparrow}-\vec{n}_{12\downarrow})^2\right)\\
 &=\sum_{\triangle}\left(\frac{\chi_{12}^b}{\chi_{13}^b}(1-\vec{n}_{13\downarrow}\cdot\vec{n}_{13\uparrow})+\frac{\chi_{13}^b}{\chi_{12}^b}(1-\vec{n}_{12\uparrow}\cdot\vec{n}_{12\downarrow})\right). 
 	\end{split}
\end{eqnarray}
There are three coordinate origins in the triangle 123, and summing over the possible contributions under the replacements $1\to 2, 2\to 3$ and $3\to 1$ with the factor $1/3$, we have
\begin{eqnarray}
	\label{discrete-form}
	\begin{split}
	&\frac{1}{3}\sum_{\Delta}\left(\frac{\chi_{12}^b}{\chi_{13}^b}(1-\vec{n}_{13\downarrow}\cdot\vec{n}_{13\uparrow})+\frac{\chi_{13}^b}{\chi_{12}^b}(1-\vec{n}_{12\uparrow}\cdot\vec{n}_{12\downarrow})\right) \\
	+&\frac{1}{3}\sum_{\Delta}\left(\frac{\chi_{23}^b}{\chi_{21}^b}(1-\vec{n}_{21\downarrow}\cdot\vec{n}_{21\uparrow})+\frac{\chi_{21}^b}{\chi_{23}^b}(1-\vec{n}_{23\uparrow}\cdot\vec{n}_{23\downarrow})\right) \\
	+&\frac{1}{3}\sum_{\Delta}\left(\frac{\chi_{21}^b}{\chi_{32}^b}(1-\vec{n}_{32\downarrow}\cdot\vec{n}_{32\uparrow})+\frac{\chi_{32}^b}{\chi_{31}^b}(1-\vec{n}_{31\uparrow}\cdot\vec{n}_{31\downarrow})\right). 
	\end{split}
\end{eqnarray}
Since $\vec{n}_{ij\uparrow}\cdot\vec{n}_{ij\downarrow}=\vec{n}_{ji\downarrow}\cdot\vec{n}_{ji\uparrow}$, we obtain
\begin{eqnarray}
	\label{discrete-H2-triangle}
	\begin{split}
		H_2=&\sum_{\Delta}H_2(\Delta),\\ 
		H_2(\Delta)=&\Gamma_{12,3}(1-\vec{n}_{12\downarrow}\cdot\vec{n}_{12\uparrow})+\Gamma_{23,1}(1-\vec{n}_{23\downarrow}\cdot\vec{n}_{23\uparrow})\\
		+&\Gamma_{31,2}(1-\vec{n}_{31\downarrow}\cdot\vec{n}_{31\uparrow}),
	\end{split}
\end{eqnarray}
where
\begin{eqnarray}
\label{Gamma-discrete-H2-triangle}
	\begin{split}
 	 \Gamma_{12,3}=&\frac{1}{3}\left(\frac{\chi_{13}^b}{\chi_{12}^b}+\frac{\chi_{23}^b}{\chi_{21}^b}\right),  \Gamma_{23,1}=\frac{1}{3}\left(\frac{\chi_{21}^b}{\chi_{23}^b}+\frac{\chi_{21}^b}{\chi_{32}^b}\right),\\
	 \Gamma_{31,2}=&\frac{1}{3}\left(\frac{\chi_{32}^b}{\chi_{31}^b}+\frac{\chi_{12}^b}{\chi_{13}^b}\right).
	\end{split}
\end{eqnarray}
The sum over triangles $\sum_{\Delta}$ can be replaced by the sum over bonds $\sum_{ij}$ such that
\begin{eqnarray}
	\label{discrete-H2-bond}
		H_2=\sum_{ij}\Gamma_{ij}^b(1-\vec{n}_{ij\downarrow}\cdot\vec{n}_{ij\uparrow}),\;
		\Gamma_{ij}^b=\frac{1}{6}\left(\frac{\chi_{il}^b}{\chi_{ij}^b}+\frac{\chi_{jl}^b}{\chi_{ji}^b}+\frac{\chi_{jk}^b}{\chi_{ji}^b}+\frac{\chi_{ik}^b}{\chi_{ij}^b}\right), 	
\end{eqnarray}
where the factor $1/3$ is replaced by  $1/6$ for $\Gamma_{ij}^b$.   The physically meaningful factor is $\kappa \Gamma_{ij}^b$, and therefore, a redefinition of $\kappa$ makes the numerical factor irrelevant, and we use $1/6$ for $\Gamma_{ij}^b$ to have the common factor with $D_{ij}^{u,v}$ corresponding to $\gamma_{ij}^{u,v}$  used in Ref. \cite{Diguet-etal-PRE2024}. The discrete forms of $H_1$, $H_{u,v}^D$ are given by
\begin{eqnarray}
	\label{discrete-H1uv-bond}
	\begin{split}
		&H_1=\sum_{ij}\Gamma_{ij}^G\ell_{ij}^2,\quad
		\Gamma_{ij}^G=\frac{1}{6}\left(\frac{\chi_{ij}^G}{\chi_{il}^G}+\frac{\chi_{ji}^G}{\chi_{jl}^G}+\frac{\chi_{ji}^G}{\chi_{jk}^G}+\frac{\chi_{ij}^G}{\chi_{ik}^G}\right), 	\\
		&H_{u}^D=\sum_{ij}D_{ij}^u(u_j-u_i)^2,\;
D_{ij}^u=\frac{1}{6}\left(\frac{\chi_{ij}^u}{\chi_{il}^u}+\frac{\chi_{ji}^u}{\chi_{jl}^u}+\frac{\chi_{ji}^u}{\chi_{jk}^u}+\frac{\chi_{ij}^u}{\chi_{ik}^u}\right), 	\\
		&H_{v}^D=\sum_{ij}D_{ij}^v(v_j-v_i)^2,\;
D_{ij}^v=\frac{1}{6}\left(\frac{\chi_{ij}^v}{\chi_{il}^v}+\frac{\chi_{ji}^v}{\chi_{jl}^v}+\frac{\chi_{ji}^v}{\chi_{jk}^v}+\frac{\chi_{ij}^v}{\chi_{ik}^v}\right). 	
	\end{split}
\end{eqnarray}

\section{Discrete RD equation} \label{App-C}
The Finsler metric version of the discrete RD equation corresponding to Eq. (\ref{discrete-t-iterations-tlattice}) is obtained by applying the variational technique to the discrete Hamiltonian
\begin{eqnarray}
	\label{Continuous-Hamil-Reaction-uv}
	H\!=\!H_u-\frac{1}{\gamma}H_v, \quad H_u=D_uH_u^D+H_u^R, \quad H_v=D_vH_v^D+H_v^R, 
\end{eqnarray}
where $H_{u,v}^D$  and $H_{u,v}^R$ are the diffusion and reaction terms, summarized as follows:
\begin{eqnarray}
	\label{Continuous-Hamil-Reaction-D}
	\begin{split}
		&H_u^D=\frac{1}{2}\int \sqrt{g}d^2x g^{ab}\frac{\partial u}{\partial x^a}\frac{\partial u}{\partial x^b}\to\sum_{ij}D_{ij}^u(u_j-u_i)^2,\\
		&H_v^D=\frac{1}{2}\int \sqrt{g}d^2xg^{ab}\frac{\partial v}{\partial x^a}\frac{\partial v}{\partial x^b}\to \sum_{ij}D_{ij}^v(v_j-v_i)^2, 
	\end{split}
\end{eqnarray}
	as described in Eq. (\ref{discrete-H1uv-bond}), and 
\begin{eqnarray}
	\label{Continuous-Hamil-Reaction-R}
	\begin{split}
		&H_u^R=-\int \sqrt{g}d^2xF\to - \sum_{i}F_i,\quad F=\frac{u^2}{2}-\frac{u^4}{4}-\frac{1}{2}uv,\\
		&H_v^R=-\gamma\int \sqrt{g}d^2xG\to -\sum_{i}G_i, \quad G=\gamma \left(\frac{1}{2}uv-\alpha \frac{v^2}{2}-\beta v\right).
	\end{split}
\end{eqnarray}
Finsler metric is assumed only in the diffusion terms, and the integral $\int \sqrt{g}d^2x$ in the reaction terms is replaced by the sum over vertices $\sum_i$ for simplicity \cite{note-8}.

The discrete variation of $H$ with respect to $u$ and $v$ is straight forward to calculate such that $\delta H\!=\!\frac{\delta H}{\delta u_i}\delta u_i\!+\!\frac{\delta H}{\delta v_i}\delta v_i$. We obtain the spatially discrete RD equation 
\begin{eqnarray}
	\label{discrete-RD-Fins}
	\begin{split}
	&\frac{\partial u_i}{\partial t}= D_u\sum_{j(i)}2D_{ij}^u (u_j-u_i) +(u_i-u_i^3-v_i),\\
	&\frac{\partial v_i}{\partial t}= D_v\sum_{j(i)}2D_{ij}^v (v_j-v_i) +\gamma (u_i-\alpha v_i-\beta)
	\end{split}
\end{eqnarray}
from $\frac{\partial u_i}{\partial t}\!=\!-\frac{\delta H}{\delta u_i}$ and  $\frac{\partial v_i}{\partial t}\!=\!-\frac{\delta H}{\delta v_i}$  to find the steady state solutions $(u_i,v_i)$,  where $\sum_{j(i)}$  denotes the sum over vertices $j$ connected to  vertex $i$.

\section{Direction-dependent interaction coefficients and energies} \label{App-D}
In this Appendix, we discuss the direction dependence of $\Gamma_{ij}^G$ and that of $\ell_{ij}^2$ of $H_1$. These quantities are defined on bond $ij$, which is not a point but has a direction described by a unit vector $\vec{e}_{ij}$ along bond $ij$ (Fig. \ref{fig-B-1}(b)). It is therefore reasonable to assume that $\Gamma_{ij}^G$ and  $\ell_{ij}^2$ are direction dependent with respect to that of $\vec{e}_{ij}$.  In addition to this direction dependence, $\Gamma_{ij}^G$ depends on the direction $\vec{\tau}$.  Therefore, it is natural to consider that  the stretching  influences  the $\vec{\tau}$ direction via $\Gamma_{ij}^G$. 

To extract the direction dependence of $\Gamma_{ij}^G$ for the Gaussian bond potential $H_1$, we define 
\begin{eqnarray}
	\label{direction-dependent-Gamma}
      \Gamma_{\mu}^G=\frac{1}{\sum_{ij}|\vec{e}_{ij}^{\;\mu}|}\sum_{ij}\Gamma_{ij}^G |\vec{e}_{ij}^{\;\mu}|, \quad  \vec{e}_{ij}^{\;\mu}=(\vec{e}_{ij}\cdot \vec{e}^{\;\mu})\vec{e}^{\;\mu}, \quad (\mu=x,y,z),
\end{eqnarray}
where $\vec{e}^{\;x}\!=\!(1,0,0)$, $\vec{e}^{\;y}\!=\!(0,1,0)$ and $\vec{e}^{\;z}\!=\!(0,0,1)$. The corresponding direction-dependent energy is defined by using 
\begin{eqnarray}
	\label{direction-dependent-H1}
		H_1^{\mu}=\frac{\sum_{ij}\Gamma_{ij}^G \ell_{ij}^2(\vec{e}_{ij}\cdot \vec{e}^{\;\mu})^2}{\Gamma_{\mu}^G}, \quad (\mu=x,y,z).
\end{eqnarray}
Using $\Gamma_{\mu}^G$, $H_1^{\mu}$ and the identity $1\!=\!(\vec{e}_{ij}\cdot \vec{e}^{\;x})^2\!+\!(\vec{e}_{ij}\cdot \vec{e}^{\;y})^2\!+\!(\vec{e}_{ij}\cdot \vec{e}^{\;z})^2$, we have 
\begin{eqnarray}
	\label{direction-dependent-coeff-energy}
		H_1=\Gamma_{x}^G H_1^{x}+\Gamma_{y}^G H_1^{y}+\Gamma_{z}^G H_1^{z},
\end{eqnarray}
and therefore, $\Gamma_{\mu}^G$ and $H_1^{\mu}$ are macroscopically meaningful for direction-dependent coefficients and energies. 

Expressions of $\Gamma_{\mu}^b$ and $H_2^{\mu}$ for the bending energy $H_2$ and of $D_{\mu}^{u,v}$ and  $H_{u,v}^{D,\mu}$ for the diffusion energies $H_{u,v}^D$ are the same as those of $\Gamma_{\mu}^G$ and $H_1^{\mu}$. Therefore, we have the decomposition of $H_u^D$ in Eq. (\ref{Continuous-Hamil-Reaction-uv}) such that 
 \begin{eqnarray}
	\label{direction-dependent-energy-u-DR}
	\begin{split}
		H_u^D&=\left(\sum_{ij}D_{ij}^u\left(u_i-u_j\right)^2 \right)
		\left((\vec{e}_{ij}\cdot \vec{e}^{\;x})^2+(\vec{e}_{ij}\cdot \vec{e}^{\;y})^2+(\vec{e}_{ij}\cdot \vec{e}^{\;z})^2\right) 	\\	
		&=D_x^u H_u^{D,x}+D_y^u H_u^{D,y}+D_z^u H_u^{D,z}
	\end{split}
\end{eqnarray}
with
 \begin{eqnarray}
	\label{direction-dependent-diff-coef}
	\begin{split}
		D_{\mu}^u&=\frac{1}{\sum_{ij}|\vec{e}_{ij}^{\;\mu}|}\sum_{ij}D_{ij}^u|\vec{e}_{ij}^{\;\mu}|, \quad (\mu=x,y,z),
	\end{split}
\end{eqnarray}
where $1\!=\!\sum_\mu(\vec{e}_{ij}\cdot \vec{e}^{\;\mu})^2, (\mu\!=\!x,y,z)$ is included in the sum in Eq. \ref{direction-dependent-energy-u-DR}). $H_v$ is also decomposed into the direction-dependent quantities as $H_u$ in Eq. (\ref{direction-dependent-energy-u-DR}).

We should emphasize that these direction-dependent quantities allow us to observe not only a non-trivial relationship between the TP direction and the mechanical properties of membranes but also how the stretching energetically modifies the stable configuration of isotropic membrane. We will further comment on this latter point:  To see the difference between the stationary configuration of the stretched membrane subsystem and that of the unstretched membrane, let us assume that the unstretched surface is in the stationary membrane configuration of the Hamiltonian $H(\vec{r},\vec{\tau})\!=\!H_1\!+\!\kappa H_2$. This stationary state is stable for all perturbations, and is therefore expected to be stable for stretching along the $x$-axis, for example, if it is small. Furthermore, this stretching is applied under the fixed boundary surface as shown in Eq. (\ref{lattice-deformation}).  For this small stretching, let us assume $\delta H\!=\!0$, and therefore, it is reasonable to expect that $\delta H_1\!=\!0$ and  $\delta H_2\!=\!0$. These conditions are almost satisfied in the region of $R_{xy}$ close to $R_{xy}\!=\!1$ as confirmed in Figs. \ref{fig-14}(c), (j) and Figs. \ref{fig-15}(c), (j). 

Under the conditions $\delta H_1\!=\!0$ and $\delta H_2\!=\!0$, since $(\vec{e}_{ij}\cdot \vec{e}^{\;z})^2$ is negligibly small compared to $(\vec{e}_{ij}\cdot \vec{e}^{\;\mu})^2, (\mu\!=\!x,y)$ in Eq. (\ref{direction-dependent-H1}), we find from Eq. (\ref{direction-dependent-coeff-energy}) that  $\delta H_1\!=\! \delta\Gamma_{x}^G H_1^{x}\!+\!\delta \Gamma_{y}^G H_1^{y}\!+\!\Gamma_{x}^G\delta H_1^{x}\!+\! \Gamma_{y}^G\delta H_1^{y}$. Due to the assumption of small stretching, $H_1^{x}\!=\!H_1^{y}$ and $\Gamma_{x}^G\!=\!\Gamma_{y}^G$ are expected up to the order of $\delta H_1^\mu$ and  $\delta\Gamma_{\mu}^G$, and therefore, we have $\delta H_1\!=\! (\delta\Gamma_{x}^G \!+\!\delta \Gamma_{y}^G) H_1^{x}\!+\! \Gamma_{x}^G (\delta H_1^{x}\!+\!\delta H_1^{y})$. Thus, $\delta H_1\!=\!0$ is satisfied for any values of $H_1^{x}(>\!0)$ and $\Gamma_{x}^G(>\!0)$ if and only if $\delta\Gamma_{x}^G \!+\!\delta \Gamma_{y}^G\!=\!0$ and $\delta H_1^{x}\!+\!\delta H_1^{y}\!=\!0$. These relations are satisfied in the two possible combinations:
 \begin{eqnarray}
	\label{two-combinations}
	\begin{split}
    &{\rm (i)\; \Gamma_x^{G}(<\Gamma_0^{G}) <\Gamma_y^{G} \Leftrightarrow H_{1}^x(>H_{1}^0) >H_{1}^y,} \\
    &{\rm (ii)\; \Gamma_x^{G}(>\Gamma_0^{G}) > \Gamma_y^{G} \Leftrightarrow H_{1}^x(>H_{1}^0) > H_{1}^y.}\\
	\end{split}
\end{eqnarray}
Note also that the symbols "$>$" and "$<$" can be exchanged in both sides of $\Leftrightarrow$ in (i) and (ii).  In both cases (i) and (ii), the energy transfer relation $ \Gamma_x^{G}H_1^x \!>\! \Gamma_y^{G}H_1^y$ ($ \Gamma_x^{G}H_1^x \!<\! \Gamma_y^{G}H_1^y$) is expected on the stretched membranes for $R_{xy}<1$ ($R_{xy}>1$). The same relations are derived for $\Gamma_\mu^b$ and $H_2^\mu$ corresponding to those in Eq. (\ref{two-combinations}). The first case (i) corresponds to the relation in Eq. (\ref{equilibrium-H12})  and is observed in the models in ${\bf R}^2$ and in the models in ${\bf R}^3$ for sufficiently large $\kappa$.  The second case (ii) is observed in model 2 in ${\bf R}^3$ for $\kappa\!=\!1$ as shown  in Figs. \ref{fig-15} (h),(i).

\section{Surface tension formula} \label{App-E}
In this Appendix, we summarize the description of surface tension in Ref. \cite{Wheater-JPA1994}. The expression in Eq. (\ref{surface-tension}) is derived for fluctuating surfaces spanning a square frame of area $A_p$ in ${\bf R}^{2}$. 

As described in Section \ref{sec:total-discrete-H}, we assume the energy unit defined by $k_BT\!=\!1$ encompassing both cases of $k_B\!=\!1$ and $T\!=\!1$, where $k_B$ and $T$ are the Boltzmann constant and the temperature. The partition function is given by the following expression:
\begin{eqnarray}
	\label{part-func}
	\begin{split}
		Z(A_p)=&\sum_\tau\int \prod_{i=1}^{N-N_{\rm fix}}d\vec{r}_i \prod_{i=1}^{N_{\rm fix}}d{r}_i\exp\left(-H\right),\\
		H=&H_1+\lambda H_\tau,
	\end{split}
\end{eqnarray}
where the symbol $A_p$ in $Z$ denotes a constraint of square boundary of a fixed area $A_p$. 
The symbol $N_{\rm fix}$ is given by
\begin{eqnarray}
	\label{effective-N}
	\begin{split}
 N_{\rm fix}=n_x+n_y,
	\end{split}
\end{eqnarray}
where $n_x$ and $n_y$ are the total number of vertices along the edges of the boundary (Fig. \ref{fig-1}(a)). The total number of boundary vertices is $2N_{\rm fix}\!-\!2$, and  the $N_{\rm fix}$ vertices at the boundary region are constrained by the periodic boundary conditions (PBCs), which dictate that they must be separated by a fixed distance from their opposite boundary vertices. Consequently, the 2-dimensional integrations $\int\prod_{i=1}^{N}d\vec{r}_i$ are effectively modified to  $\int\prod_{i=1}^{N-N_{\rm fix}}d\vec{r}_i\prod_{i=1}^{N_{\rm fix}}d{r}_i$, where $\int\prod_{i=1}^{N-N_{\rm fix}}d\vec{r}_i$ and $\int\prod_{i=1}^{N_{\rm fix}}d{r}_i$ are  2-dimensional and 1-dimensional integrations, respectively. The latter part $\int\prod_{i=1}^{N_{\rm fix}}d{r}_i$ denotes the reduced integrations for the vertices at the boundaries constrained by PBCs such that $\int\prod_{i=1}^{n_y}d{r}_i\!=\!\int\prod_{i=1}^{n_y}dx_i$ and $\int\prod_{i=n_y\!+\!1}^{N_{\rm fix}}d{r}_i\!=\!\int\prod_{i=1}^{n_x}dy_i$. These reduced integrations are 2-dimensional such that $\int\prod_{i=1}^{N_{\rm fix}}d\vec{r}_i$ in the case of the models in ${\bf R}^3$.

The integration variable can be transformed from $\vec{r}$ to $\alpha \vec{r}, (\alpha\!>\!0)$, because the integration is generally independent of the variable. This implies that $Z(\vec{r};A_p)\!=\!Z(\alpha \vec{r};A_p(\alpha))$. From this, we obtain the relation $\left. \frac{1}{Z(\alpha \vec{r})}\frac{d Z(\alpha \vec{r})}{d \alpha}\right|_{\alpha=1}\!=\!0$. Since the scaled $Z$ is given by
\begin{eqnarray}
	\begin{split}
		\label{scaled part-func}
		&Z(\alpha \vec{r};A_p(\alpha))\\
		=&
		\alpha^{2({N-N_{\rm fix})+N_{\rm fix}}}\sum_\tau\int \prod_{i=1}^{N-N_{\rm fix}}d\vec{r}_i \prod_{i=1}^{N_{\rm fix}}d{r}_i\exp\left(-H(\alpha \vec{r})\right),\\
		&H(\alpha \vec{r})=\alpha^2 H_1+\lambda H_\tau,
	\end{split}
\end{eqnarray}
we have
\begin{eqnarray}
	\label{derivation-of-Z}
	\begin{split}
		&\frac{dZ(\alpha \vec{r};A_p(\alpha))}{d\alpha}
		=2{N-N_{\rm fix}}\alpha^{2N-N_{\rm fix}-1} Z(\alpha \vec{r})
		+ \frac{\partial Z}{\partial A_p}\frac{\partial A_p}{\partial \alpha}\\
		&-2\alpha \sum_\tau\int \prod_{i=1}^{N}d\vec{r}_i \prod_{i=1}^{N_{\rm fix}}d{r}_i H_1\exp\left(-H(\alpha \vec{r})\right).
	\end{split}
\end{eqnarray}
Note that the potential $U_V$, defined in Eq. (\ref{discrete-Hamiltonian}),  is not included in $H$ of Eq. (\ref{part-func}), since $U_V$ remains unchanged under the expansion $\vec{r}\!\to\! \alpha \vec{r}$ if the lattice spacing $a$ or the numbers in $\ell_{\rm min}$ and $\ell_{\rm max}$ are changed according to $\vec{r}\!\to\! \alpha \vec{r}$. The lattice spacing $a$ is defined by the side length of the regular triangle of the initial configuration (Fig. \ref{fig-1}(a)), and it can be expanded according to $\vec{r}\!\to\! \alpha \vec{r}$. In the simulations, both $a$ and the numbers $\ell_{\rm min}$ and $\ell_{\rm max}$ are held constant, independent of $R_{xy}$. However, as shown in Eq. (\ref{lattice-deformation}), the surface area remains constant and is independent of the shape deformation caused by the variation of $R_{xy}$. Furthermore, the range of $U_V$ defined by the $\ell_{\rm min}$ and $\ell_{\rm max}$ is sufficiently extensive  that  the neglect of $U_V$ is not expected to have any significant effect on the results.
	
The fixed boundary area $A_p$ implies that $A_p(\alpha)\!=\!\alpha^{-2}A_p$ under $\vec{r}\!\to\! \alpha \vec{r}$, and therefore, we obtain $\frac{\partial Z}{\partial A_p}\frac{\partial A_p}{\partial \alpha}\!=\!-2\alpha^{-3}A_p \frac{\partial Z}{\partial A_p}$. 
To calculate $\frac{\partial Z}{\partial A_p}$, we assume that the free energy of membrane is given by
\begin{eqnarray}
	\label{free-energy-F}
	F=\int^{A_p}\sigma(A) dA,\quad Z=\exp\left(-F\right)
\end{eqnarray}
using the surface tension $\sigma(A)$, and we obtain 

	\begin{eqnarray}
		\frac{\partial Z}{\partial A_p}=-\frac{\partial F}{\partial A_p}Z=-\left(\frac{\partial }{\partial A_p}\int^{A_p}\sigma(A) dA\right) Z=-\sigma(A_p) Z.
	\end{eqnarray}
Note that the initial area $A_0$ should be incorporated into the integral of $F$ in Eq. (\ref{free-energy-F}) to ensure the well-defined nature of  $F$. However, as mentioned in the main text, $A_0$ is  excluded from the integral  due to the utilisation of solely its derivative $\partial F/\partial A_p$, which is independent of $A_0$.

In the case of surfaces in ${\bf R}^2$, we have 
\begin{eqnarray}
	\label{derivation-of-Z-formula-2D}
	\begin{split}
		&0=\left. \frac{1}{Z}\frac{dZ(\alpha \vec{r};A_p(\alpha))}{d\alpha}\right|_{\alpha=1}
=2N-N_{\rm fix}+ 2\sigma A_p -2\langle H_1\rangle\\
&\Leftrightarrow 
	\sigma=\frac{H_1}{A_p}-\frac{N-N_{\rm fix}/2}{A_p}\\
	&\quad=\frac{1}{A_p}\left(\sum_{ij} \Gamma_{ij}^G\ell_{ij}^2-\frac{1}{2}\left(2N-N_{\rm fix}\right)\right),\quad ({\bf R}^2).
	\end{split}
\end{eqnarray}

It is important to note that the free energy $F$ in Eq. (\ref{free-energy-F}) is derived under the assumption that the surface tension $\sigma$ depends only on the projected area $A$ and is independent not only of the surface shape but also of the internal structure. For this reason,  $\sigma$ has no information on whether the triangulated surfaces is fluid or fixed at this stage.  On the other hand, when this $\sigma$ is employed in the second term on the right hand side of  the first line of Eq. (\ref{derivation-of-Z-formula-2D}), the final expression for $\sigma$ in Eq. (\ref{derivation-of-Z-formula-2D}) is obtained. This expression of $\sigma$ depends on the microscopic structure of the membranes via $H_1$, whereby $\Gamma_{ij}^G(\tau)$ varies depending on the shape of surface $R_{xy}$ even when the projected  $A_p$ is held constant. For this reason, the isotropic $\sigma$ is  only expected to be correct for $R_{xy}\!=\!1$.  The directional surface tension is therefore a useful concept, which will be introduced below.

The isotropic surface tension $\sigma$ for the models in ${\bf R}^3$ can be obtained by the expansion $\alpha \vec{r}\!=\!(\alpha x, \alpha y, \alpha z)$.  In this case, we obtain $0\!=\!3N\!-\!N_{\rm fix}\!+\! 2\sigma A_p \!-\!2\langle H_1\rangle$ corresponding to the first line of Eq. (\ref{derivation-of-Z-formula-2D}), and therefore,
\begin{eqnarray}
\label{derivation-of-Z-formula-3D-iso}
	\sigma=\frac{1}{A_p}\left(H_1-\frac{1}{2}\left(3N-N_{\rm fix}\right)\right),\quad ({\bf R}^3).
\end{eqnarray}
This surface tension is correct only when the pressure difference between the upside and downside of the membrane is neglected,  along with the thickness.  However, when the thickness is not negligible, it is imperative to utilise three-dimensional stress instead  of surface tension.  In this manuscript, we assume that the membranes are two-dimensional sheets in which the thickness is neglected, as mentioned in the first of Section \ref{Sec:lattices}. The relation between $\sigma$ and $\sigma^x$, which is introduced below, is discussed in the supplementary material (2).


\section{Direction-dependent surface tension} \label{App-F}
Now we consider the new integration variable  $(\alpha x,y)$ in the partition function $Z$ for the models in ${\bf R}^2$, corresponding to $Z$ in Eq. (\ref{part-func}), instead of $\vec{r}\!=\!(x,y)$ using an expansion vector $\vec{\alpha}\!=\!(\alpha,1)$ with a number $\alpha\!>\!0$. We simply express the new variables by $\vec{r}_{\vec{\alpha}}\!=\!(\alpha x,y)$, which is different from ${\alpha}\vec{r}$ for the isotropic case in Appendix \ref{App-E}. We focus on the case of the expansion along the $x$ direction. 

In the case of $\vec{r}_{\vec{\alpha}}\!=\!(\alpha x,y)$, $H_1(\vec{r})\!=\!\sum_{ij}\Gamma_{ij}^G\ell_{ij}^2$ can be approximated to  
	\begin{eqnarray}
		\label{x-scaled-H1}
		\begin{split}
			H_1(\vec{r}_{\vec{\alpha}})=	\sum_{ij}\Gamma_{ij}^G\ell_{ij}^2\left(\alpha^2(\vec{e}_{ij}\cdot \vec{e}^{\;x})^2+(\vec{e}_{ij}\cdot \vec{e}^{\;y})^2\right).
		\end{split}
	\end{eqnarray}
because $\ell_{ij}^2\!=\!(\vec{r}_j\!-\!\vec{r}_i)^2$ changes as follows:
	\begin{eqnarray}
		\begin{split}
			\ell_{ij}^2= &(x_j-x_i)^2+(y_j-y_i)^2 \\
			\to &\alpha^2(x_j-x_i)^2+(y_j-y_i)^2 \\
			=&\left(\alpha^2(\vec{\ell}_{ij}\cdot \vec{e}^{\;x})^2+(\vec{\ell}_{ij}\cdot \vec{e}^{\;y})^2\right).
		\end{split}
	\end{eqnarray}
We should note that the variation of $\Gamma_{ij}^G$ under  $(\alpha x,y)$ is negligible in comparison to that of $\ell_{ij}^2$. This assumption is reasonable   for the first approximation because the lattice average of $\Gamma_{ij}^G$ is almost independent of the lattice deformation by $R_{xy}$ (see Fig. 2 of the supplementary material (2)),  though $(\alpha x,y)$ is not always identical to the lattice deformation by $R_{xy}$ and the $\vec{\tau}$ variation is also reflected in the data plotted in Fig. 2 of the supplementary material (2).  The right hand side of $H_1(\vec{r}_{\vec{\alpha}})$ in Eq. (\ref{x-scaled-H1}) can also be written as
	\begin{eqnarray}
		\label{x-scaled-H1-direc-dep}
		\begin{split}
			H_1(\vec{r}_{\vec{\alpha}})= \alpha^2	\Gamma_x^G H_1^x+\Gamma_y^G H_1^y
		\end{split}
	\end{eqnarray}
 by	using $H_1^\mu$ in Eq. (\ref{direction-dependent-H1}) and the decomposition in Eq. (\ref{direction-dependent-coeff-energy}) for the case of ${\bf R}^2$.

 Due to the identity for the partition function $Z(\vec{r}; A_p(\vec{r}))\!=\!Z(\vec{r}_{\vec{\alpha}}; A_p(\vec{r}_{\vec{\alpha}}))$, we obtain  the relation $\left. \frac{1}{Z(\vec{r}_{\vec{\alpha}})}\frac{d Z(\vec{r}_{\vec{\alpha}})}{d \alpha}\right|_{\alpha=1}\!=\!0$. 
 Since the scaled $Z$ is given by
\begin{eqnarray}
	\begin{split}
		\label{scaled part-func-dir}
		Z(\vec{r}_{\vec{\alpha}};A_p(\vec{r}_{\vec{\alpha}}))=&
		\alpha^{N-N_{\rm fix}+n_x}\sum_\tau\int \prod_{i=1}^{N-N_{\rm fix}}d\vec{r}_i \prod_{i=1}^{N_{\rm fix}}d{r}_i\exp\left(-H(\vec{r}_{\vec{\alpha}})\right),\\
		H(\vec{r}_{\vec{\alpha}})=&H_1(\vec{r}_{\vec{\alpha}})+\lambda H_\tau\\
		=& \left(\alpha^2	\Gamma_x^G H_1^x+\Gamma_y^G H_1^y\right)+\lambda H_\tau,
	\end{split}
\end{eqnarray}
we have
\begin{eqnarray}
	\label{derivation-of-Z-dir}
	\begin{split}
		&\frac{dZ(\vec{r}_{\vec{\alpha}};A_p(\vec{r}_{\vec{\alpha}}))}{d\alpha}
		=(N-n_y)\alpha^{N-n_y-1} Z(\vec{r}_{\vec{\alpha}})
		+ \frac{\partial Z}{\partial A_p}\frac{\partial A_p}{\partial \alpha}\\
		&-2\alpha \sum_\tau\int \prod_{i=1}^{N-N_{\rm fix}}d\vec{r}_i\prod_{i=1}^{N_{\rm fix}}d{r}_i \,\Gamma_x^GH_1^x\,\exp\left(-H(\vec{r}_{\vec{\alpha}})\right).
	\end{split}
\end{eqnarray}
 Under the expansion defined by  $\vec{r}_{\vec{\alpha}}\!=\!(\alpha x,y,z)$, the fixed frame of area $A_p$ should be scaled only in the $x$ direction such that $A_p(\vec{r}_{\vec{\alpha}})\!=\!\alpha^{-1}A_p(\vec{r})$, which corresponds to $A_p(\alpha\vec{r})\!=\!\alpha^{-2}A_p(\vec{r})$ in the isotropic case in Appendix \ref{App-E}. Therefore, we obtain $\frac{\partial Z}{\partial A_p}\frac{\partial A_p}{\partial \alpha}\!=\!-\alpha^{-2}A_p \frac{\partial Z}{\partial A_p}$.

 To calculate $\frac{\partial Z}{\partial A_p}$, we assume that the free energy of membrane is given by
\begin{eqnarray}
	\label{free-energy-F-dir}
	F=\int^{A_p}\sigma^x(A)dA,\quad Z=\exp\left(-F\right)
\end{eqnarray}
using the surface tension $\sigma^x(A)$, and we obtain 
\begin{eqnarray}
		\frac{\partial Z}{\partial A_p}=-\frac{\partial F}{\partial A_p}Z=-\left(\frac{\partial }{\partial A_p}\int^{A_p}\sigma^x(A) dA\right) Z=-\sigma^x(A_p) Z.
	\end{eqnarray}

 Thus, we have
\begin{eqnarray}
	\label{derivation-of-Z-formula-2D-dir}
	\begin{split}
		0=&\left. \frac{1}{Z}\frac{dZ(\vec{r}_{\vec{\alpha}};A_p(\vec{r}_{\vec{\alpha}}))}{d\alpha}\right|_{\alpha=1}
		=N-n_y+ \sigma^x A_p -2 \Gamma_x^G H_1^x\\
		\Leftrightarrow \sigma^x=&\frac{2\Gamma_x^GH_1^x}{A_p}-\frac{N-n_y}{A_p} \\
		=&\frac{1}{A_p}\left(2\sum_{ij}\Gamma_{ij}^G \ell_{ij}^2(\vec{e}_{ij}\cdot \vec{e}^{\;x})^2-(N-n_y)\right),\quad ({\bf R}^2).
	\end{split}
\end{eqnarray}
We should note that $\sigma^x$ is comparable to $\sigma$ in Eq. (\ref{derivation-of-Z-formula-2D}) on the membranes in ${\bf R}^2$ at $R_{xy}\!=\!1$, where $\Gamma_x^GH_1^x\!\simeq\!H_1/2$ is expected. 
The expressions of $\sigma^y$ can also be obtained with the same procedure described above using the expansion vector $\vec{\alpha}\!=\!(1,\alpha)$.

In the case of surfaces in ${\bf R}^3$, the bending energy $H_2$ also varies due to  the expansion $\vec{r}_{\vec{\alpha}}\!=\!(\alpha x,y,z)$, and therefore, we have $N\!-\!n_y\!+\!\sigma^x A_p\!-\!2\Gamma_x^GH_1^x\!-\!\kappa \frac{\partial H_2}{\partial \alpha}|_{\alpha=1}\!=\!0$ corresponding to the first expression of Eq. (\ref{derivation-of-Z-formula-2D-dir}). Therefore, we obtain
\begin{eqnarray}
		\label{derivation-of-Z-formula-3D-dir}
	\sigma^x=\frac{2\Gamma_x^GH_1^x}{A_p}+\frac{1}{A_p}\kappa \left.\frac{\partial H_2(\alpha x,y,z)}{\partial \alpha} \right|_{\alpha=1}-\frac{N-n_y}{A_p},\quad ({\bf R}^3).
\end{eqnarray}
It is straight forward to calculate $\left.\frac{\partial H_2(\alpha x,y,z)}{\partial \alpha} \right|_{\alpha=1}$ by expressing the normal vectors with the new variable $(\alpha x,y,z)$ under the assumption that the variation of intensive part $\Gamma_{ij}^b$ is small compared to that of the extensive part.

We evaluate the approximate order of $\sigma^x$ in Eq. (\ref{derivation-of-Z-formula-3D-dir}) in the physical unit $({\rm N/m})$. In the physical unit, the projected area $A_p$ is replaced by $A_p a^2$ with the lattice spacing $a ({\rm m})$ and $2\Gamma_x^GH_1^x\!+\!\kappa \left.\frac{\partial H_2(\alpha x,y,z)}{\partial \alpha} \right|_{\alpha=1}\!-\!(N\!-\!n_y)$ is multiplied by $k_BT ({\rm Nm})$, and we have 
\begin{eqnarray}
	\label{surface-tension-phys}
	\sigma_{\rm phy}^x=\sigma^x \frac{k_BT}{a^2} \left({\rm N/m}\right) 
	=  4\times 10^{-21}\frac{\sigma^x}{a^2}  \left({\rm N/m}\right),
\end{eqnarray}
where $k_B\!=\! 1.38\times 10^{-23} {\rm J/K}$, and the room temperature $T\!=\!300 {\rm K}$ is assumed. The lattice spacing $a$ should be larger than the monomer size (typically $\sim 10^{-9} {\rm m}$), as described in Section \ref{sec:total-discrete-H}.  By assuming that $a\!\simeq\!10^{-8} {\rm m}$ and using the simulation data $\sigma^x\!\simeq\!1$ in Figs. \ref{fig-10}(c)--(e), we have $\sigma_{\rm phy}^x\!\simeq\! 10^{-5} {\rm N/m}$. In the case that $a\!\simeq\!10^{-7} {\rm m}$,  we have $\sigma_{\rm phy}^x\!\simeq\! 10^{-7} {\rm N/m}$.

\balance

\renewcommand\refname{Notes and references}


\begin{thebibliography}{9}
\bibitem{note-1}
The diffusions, described by $D_u\Laplace u$ and $D_v\Laplace v$, are also considered to be caused by chemical reactions. However, the standard terminology is employed, whereby "reaction" is used for $f$ and $g$, and  "diffusion" is used for $D_u\Laplace u$ and $D_v\Laplace v$.
\bibitem{note-3} 
The coefficient $\Gamma_{ij}^G$, along with $\Gamma_{ij}^b$ and $D_{ij}^{u,v}$, is dependent on the lattice structure $\vec{r}$. However, we utilise the notation $\Gamma_{ij}^G(\tau)$ to underscore the non-trivial dependence on $\vec{\tau}$.
\bibitem{note-4} 
The definition of $H_\tau$ is modified from $H_\tau\!=\!\frac{1}{2}\sum_{ij}\left(1\!-\!3\left(\vec{\tau}_i\cdot \vec{\tau}_j\right)^2\right)$ for the liquid crystals to $H_\tau\!=\!\frac{3}{2}\sum_{ij}\left(1\!-\!\left(\vec{\tau}_i\cdot \vec{\tau}_j\right)^2\right)$, which satisfies $H_\tau\!\geq\!0$.
\bibitem{note-5} 
This area $A_p$ is assumed to vary for the evaluation of the surface tension, with the variation by $R_{xy}$ being fixed.
\bibitem{note-7} 
In the case that $\vec{e}_{ij}\!\parallel\!\vec{e}^{\,x}$, specifically in the case of $\vec{e}_{ij}$ and $\vec{e}^{\,x}$ being oriented in the same direction,  we have $R_{ij}^G|\vec{e}_{ij}^{\,x}|\!=\!R_{ij}^G(x)\simeq 0$ as illustrated in Fig. \ref{fig-12}(b);  $R_{ij}^G$ is denoted by $R_{ij}^G(x)$. While in the case  that $\vec{e}_{ij}\!\parallel\!\vec{e}^{\,y}$,  we have $R_{ij}^G(y)|\vec{e}_{ij}^{\,y}|\!=\!R_{ij}^G(y)\!>\!0$ as illustrated in Fig. \ref{fig-12}(b). Thus, we have $R_{ij}^G(x)<R_{ij}^G(y)$, which satisfies the first relation in Eq. (\ref{discussion-summary}). Thus, we confirm that this microscopic property is accurately  reflected in the macroscopic $\Gamma_\mu^G$ in Eq. (\ref{effective-surface-tension}).
\bibitem{note-8} 
In this paper, as well as in \cite{Diguet-etal-PRE2024}, the Finsler metric is assumed solely in the diffusion terms for the anisotropic diffusion.  It is also possible to assume the Finsler metric in the reaction terms; however, the results remain almost unchanged.
\end{thebibliography}

\end{document}